\documentclass[%
reprint,
superscriptaddress,
amsmath,amssymb,
aps,
prx,
]{revtex4-2}
\usepackage{graphicx}
\usepackage{dcolumn}
\usepackage{bm}
\usepackage{chemformula}
\usepackage{hyperref}
\usepackage{tabularx}
\usepackage{amsmath}
\usepackage{float}

\begin{document}	
 \title{Anomalous Proximitized Transport in Metal/Quantum Magnet Heterostructure \ch{Bi_2Ir_2O_7}/\ch{Yb_2Ti_2O_7}}
    \author{Chengkun Xing}
	\affiliation{Department of Physics and Astronomy, University of Tennessee, Knoxville, TN 37996, USA}
        \author{Shu Zhang}
	\affiliation{Max-Planck-Institut fur Physik komplexer Systeme, Nothnitzer Straße 38, 01187 Dresden, Germany}
	\author{Weiliang Yao}
	\affiliation{Department of Physics and Astronomy, University of Tennessee, Knoxville, TN 37996, USA}
        \author{Dapeng Cui}
	\affiliation{Department of Physics and Astronomy, University of Tennessee, Knoxville, TN 37996, USA}
	\author{Qing Huang}
	\affiliation{Department of Physics and Astronomy, University of Tennessee, Knoxville, TN 37996, USA}
	\author{Junyi Yang}
	\affiliation{Department of Physics and Astronomy, University of Tennessee, Knoxville, TN 37996, USA}
	\author{Shashi Pandey}
	\affiliation{Department of Physics and Astronomy, University of Tennessee, Knoxville, TN 37996, USA}
        \author{Dongliang Gong}
	\affiliation{Department of Physics and Astronomy, University of Tennessee, Knoxville, TN 37996, USA}
	\author{Lukas Horák}
	\affiliation{Department of Condensed Matter Physics, Faculty of Mathematics and Physics, Charles University, Prague, 12116, Czech Republic}
	\author{Yan Xin}
	\affiliation{National High Magnetic Field Laboratory, Florida State University, Tallahassee, FL 32310, USA}
	\author{Eun Sang Choi}
	\affiliation{National High Magnetic Field Laboratory, Florida State University, Tallahassee, FL 32310, USA}
 	\author{Yang Zhang}
	\email{yangzhang@utk.edu}
	\affiliation{Department of Physics and Astronomy, University of Tennessee, Knoxville, TN 37996, USA}
 	\affiliation{Min H. Kao Department of Electrical Engineering and Computer Science, University of Tennessee, Knoxville, TN 37996, USA}
	\author{Haidong Zhou}
	\email{hzhou10@utk.edu}
	\affiliation{Department of Physics and Astronomy, University of Tennessee, Knoxville, TN 37996, USA}
	\author{Jian Liu}
	\email{jianliu@utk.edu}
	\affiliation{Department of Physics and Astronomy, University of Tennessee, Knoxville, TN 37996, USA}
	
	\date{\today}%
	
	\begin{abstract}
		{Fluctuations of quantum spins play a crucial role in the emergence of exotic magnetic phases and excitations. The lack of the charge degree of freedom in insulating quantum magnets, however, precludes such fluctuations from mediating electronic transport. Here we show that the quantum fluctuations of a localized frustrated magnet induce strong proximitized charge transport of the conduction electrons in a synthetic heterostructure comprising an epitaxial \ch{Bi_2Ir_2O_7} ultrathin film on the single crystal of \ch{Yb_2Ti_2O_7}. The proximity effects are evidenced by the scaling behavior of the \ch{Bi_2Ir_2O_7} resistance in correspondence with the dynamic scaling of the dynamic spin correlation function of \ch{Yb_2Ti_2O_7}, which is a result of quantum fluctuations near a multi-phase quantum critical point. The proximitized transport in \ch{Bi_2Ir_2O_7} can be effectively tuned by magnetic field through suppressing the quantum spin fluctuations as well as inducing transitions via magnetic anisotropy in \ch{Yb_2Ti_2O_7}. Our work establishes a new pathway for harnessing quantum spin fluctuations in magnetic insulators with electric transport, offering exciting prospects for potential applications in the realm of quantum spintronics.}
	\end{abstract}
	\maketitle
 

\section*{Introduction}
Spin-electronics roots in the dependence of electronic transport on the magnetization \cite{wolf2001spintronics,hirohata2014magnetic}. This concept has been extended well beyond ferromagnets to antiferromagnets, ferrimagnets, and other magnets where the magnetic order can be described as a classical vector \cite{baltz2018antiferromagnetic,jungwirth2016antiferromagnetic,gomonay2014spintronics}.Electronic transport can be controlled in such magnetic systems by reorienting the moments, which however must overcome energy barriers and inevitably cause power dissipation. Quantum mechanics, on the other hand, provides an attractive alternative for one spin state to transit to another spin state through quantum tunneling. Such quantum effect is one of the salient features of quantum magnets, where the magnetic state is a superposition of multiple distinct collective spin configurations due to quantum spin fluctuation (QSF) \cite{sachdev1999quantum,balents2010spin,zhou2017quantum,broholm2020quantum}. 
Static magnetic orders are thus often strongly suppressed in quantum magnets, and the conventional quasiparticle picture \cite{stone2006quasiparticle} may break down in describing the low-energy excitations due to the absence of sharp spin waves. 
Indeed, exotic QSF-driven excitations are believed to be highly correlated and entangled, holding promises for quantum computation \cite{nayak2008non,savary2016quantum,zhou2017quantum,broholm2020quantum,balents2010spin,taillefumier2017competing,takahashi2019spin,aasen2020electrical}, with spinon and Majornana fermion being two well-known examples that have been intensively studied in quantum magnets \cite{kitaev2006anyons,banerjee2017proximate,kasahara2018majorana,yokoi2021quantum,trebst2022kitaev}.

Despite the great technological potential \cite{tokura2017emergent}, it is not yet clear how to exploit such many-body magnetic quantum effects to control electronic transport. A key reason is that the spin degree of freedom in most quantum magnets originates from strongly localized electronic states and they interact without free electrons \cite{balents2010spin,norman2016colloquium,broholm2020quantum,wen2019,clark2021}. In fact, many intriguing quantum magnets are good insulators. \ch{Yb_2Ti_2O_7} (YbTO) is an excellent example of such insulating quantum magnets \cite{ross2009,thompson2011,yaouanc2011,Chang2012,DOrtenzio2013,Lhotel2014,Thompson2017}, which is currently under extensive studies due to the proposed quantum spin ice state and multiphase competition \cite{ross2011,Robert2015,Gaudet2016,Pan2016,PecanhaAntonio2017}. Its ferromagnetic (FM) ground state is strongly suppressed \cite{Arpino2017} with very broad dispersive spin waves due to a significant mixture from antiferromagnetic (AFM) configurations [Fig. \ref{fig1}] alongside an intense flat mode near the spin-wave gap \cite{scheie2017reentrant,Scheie2020a,petit2020way,scheie2022dynamical}.
This intriguing quantum dynamics arises from the strong QSF near the FM-AFM phase boundary \cite{Scheie2020a,petit2020way}, and persists at finite temperatures even above $T_{c}$ \cite{scheie2017reentrant,saubert2020}. If an effective coupling to charge carriers can be introduced while preserving the quantum dynamics, one may exploit such QSF of local moments to mediate electronic transport for better understanding of the quantum magnetism and enabling new spin-electronic functionality.

Here we demonstrate proximitized electric transport for investigating and harnessing the quantum effects of YbTO crystals by forming an epitaxial heterostructure [Fig. \ref{fig1}] with ultrathin film of nonmagnetic metal \ch{Bi_2Ir_2O_7} (BIO). The feasibility of this hetero-epitaxial synthesis \cite{zhang2023anomalous} was recently demonstrated between BIO and \ch{Dy_2Ti_2O_7} (DTO) \cite{bramwell2001spin,siddharthan1999ising,ramirez1999zero,castelnovo2008magnetic,tabata2006kagome,jaubert2013topological}. In contrast to the classical spin ice physics of DTO, we find that the interfacial impact on the electronic transport of the BIO thin film is drastically enhanced by YbTO due to the QSF of YbTO, which is highlighted by an anomalous scaling of the BIO resistivity with temperature and magnetic field that reflects the unconventional spin-wave dynamics of YbTO. Our findings provide compelling evidence for a distinct avenue to access and explore quantum magnetism through proximitized electric transport. The realization of transport control of itinerant electrons through localized quantum spin degrees of freedom in the insulator could enable potential applications in spintronic devices.

\section*{Results}
\subsection{Synthesis}
\begin{figure}[t]
    \centering
    \includegraphics[clip,width=8cm]{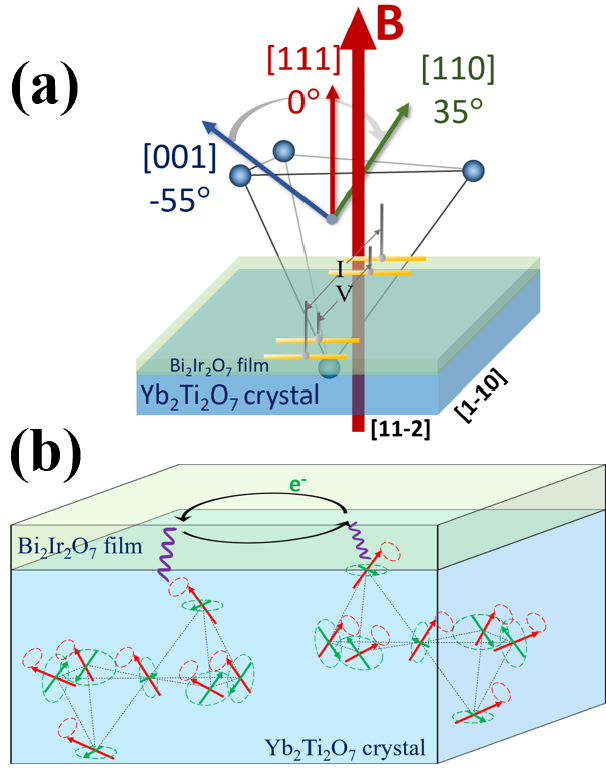}
    \caption{(a) Schematic diagram of the interface of BIO/YbTO heterostructure and the measurement configuration. (b) Schematic diagram of the scattering process on the interface of BIO/YbTO heterostructure. The tetrahedra illustrate the coexistence of FM (red arrows with freedom to rotate around the local [001] axis) and AFM (green arrows with freedom to totate in the local (111) plane) phase of YbTO crystal structure.}\label{fig1}
\end{figure}

To couple the QSF of localized spins with charge carriers, we devised and synthesized the epitaxial heterostructure of \ch{Bi_2Ir_2O_7}/\ch{Yb_2Ti_2O_7} (BIO/YbTO) [Fig. \ref{fig1}] by depositing a nominally 4 nm-thick BIO thin film onto a (111)-oriented stoichiometric YbTO single crystal substrate. BIO is an ideal carrier provider since it is a nonmagnetic metal with a pyrochlore structure~\cite{chu2019} compatible with YbTO. Since the properties of YbTO is believed to be sensitive to disorder, we prepared stoichiometric YbTO single crystal with a modified traveling-solvent floating zone method as recently reported \cite{Arpino2017}. The lattice parameter of our stoichiometric YbTO single crystal is 10.03377(5) $\rm{\AA}$ [Fig. S1(b)] \cite{supplementary}, consistent with the previous report \cite{Arpino2017}. The epitaxy of the heterostructure was confirmed by cross-section transmission electron microscope (TEM) [Fig. S2(b)] \cite{supplementary} with energy dispersive spectroscopy [Fig. S3] \cite{supplementary}. X-ray diffraction [Fig. S2(c)] \cite{supplementary} also confirmed the single-phase epitaxial growth. The actual thickness of the BIO film to be discussed below was measured to be 3.9 nm [Fig. S2(e)] \cite{supplementary} by X-ray reflectivity. 
\subsection{Proximitized Transport due to FM-AFM Competition}

\begin{figure*}[t]
    \centering
    \includegraphics[width=1\textwidth]{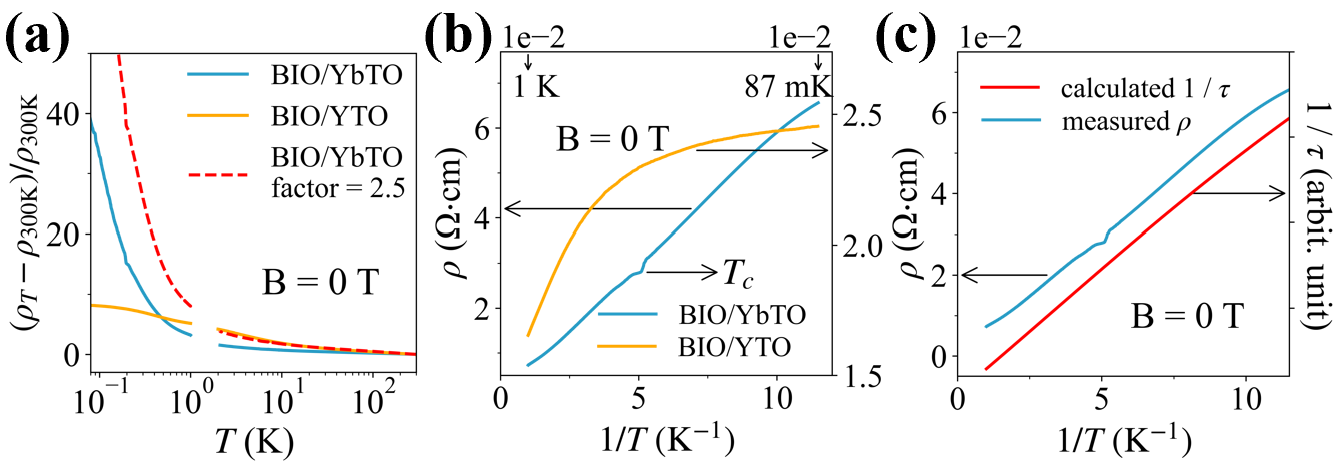}
    \caption{(a) Temperature dependence of the normalized resistivity for the BIO/YbTO (blue) and BIO/YTO (orange) heterostrutures from 300 K to 87 mK at zero magnetic field. The temperature gap between 1 to 2 K is due to the fact that two instruments were used to measure the high and low temperature resistivity, respectively. The red dashed line is obtained by multiplying the blue line with a factor of 2.5 to illustrate the similar temperature dependence of the two samples at high temperatures and their drastic difference at low temperature. In other words, it is possible to scale the two samples at high temperature region but not the low temperature region.
    (b) The zero-field resistivity plotted as a function of $1/T$ in the low temperature region. While the BIO/YbTO curve shows a linear behavior down to about 100 mK, the BIO/YTO curve is clearly sub-linear. The kink position on BIO/YbTO curve corresponds to the FM transition temperature of YbTO at zero field. (c) The inverse effective relaxation time (red line) calculated from  Eq.~(\ref{eq:tau}) shows a same $1/T$ dependence as the resistivity does.}\label{fig2}
\end{figure*}

The influence of YbTO magnetism becomes evident when comparing the resistivity of the BIO/YbTO heterostructure with a reference sample where the BIO film was grown on a nonmagnetic \ch{Y_2Ti_2O_7} (YTO) single crystal substrate [Fig. \ref{fig2}(a)]. In the high-temperature regime, both samples exhibit a similar trend as a function of temperature. The resistivity increases slowly with decreasing temperature \cite{qi2012strong} due to weak localization and/or disorder-enhanced electron-electron interaction in two-dimensions \cite{chu2019}, especially at such reduce thickness. Nonetheless, the resistivity curves above 2 K can be mapped to each other by a scaling factor between BIO/YbTO and BIO/YTO heterostructures. However, in the sub-Kelvin regime, thermal fluctuations are suppressed, the impact of quantum magnetism takes over, and notable distinctions arise. Specifically, the resistivity of BIO/YbTO experiences a sharp increase at low temperatures, following a $1/T$ scaling [Fig. \ref{fig2}(b)], i.e., $\rho \propto 1/T$. Remarkably, this scaling behavior persists even below the FM transition temperature $T_{c}$ of YbTO ($\sim$ 224 mK [Fig. S4(a)] \cite{supplementary}). In contrast, the resistivity of BIO/YTO remains varying gradually, and the $1/T$ scaling is clearly absent. Similarly, the $1/T$ scaling was not observed in the BIO/DTO heterostructure [see Fig. S5] \cite{supplementary}.

  
We find this scaling behavior highly unusual. In the simplest Drude picture, the resistivity is inversely proportional to an effective relaxation time $\tau$. If the proximity effect arises from scattering processes due to the exchange interactions between the spins of the charge carriers and the localized spins in the magnetic insulator like YbTO, the resultant electron relaxation time will have a temperature dependence governed by the spin fluctuations in the magnetic insulator~\cite{balberg1977critical}:
 \begin{equation}\label{eq:tau}
    \frac{1}{\tau (T)} \propto \int \mathrm{d} \omega \int \mathrm{d}^2 \mathbf{q} \; \mathcal{S}(\mathbf{q}, \omega, T) \Phi (\mathbf{q}),
\end{equation}
where $\mathcal{S}(\mathbf{q}, \omega, T)$ is the dynamical spin correlation functions in YbTO, and $\Phi (\mathbf{q})$ is a form factor that accounts for the phase space of scattering determined by the Fermi surface structure of BIO. 
In the temperature range of interests, one can assume that other scattering channels, such as electron-electron, electron-phonon, and impurity scatterings, have a much weaker temperature dependence.
The observed $1/T$ scaling behavior then clearly contradicts what one would expect based a conventional spin-wave picture for an FM order, where the thermal occupation of bosonic spin waves increases as temperature increases and yields a shorter relaxation time, leading to a positive contribution to the resistivity.
Instead, a previous theoretical study has shown that the $1/T$ behavior of itinerant frustrated magnets can be attributed to quantum fluctuations in a liquid-like spin state below the Curie-Weiss temperature~\cite{Batista2016prl}.

To evaluate such a mechanism, we refer to the dynamical structure factor $\mathcal{S}(\mathbf{q}, \omega, T)$ reported by a recent inelastic neutron scattering experiment~\cite{scheie2022dynamical} on YbTO in the low-energy regime of tens of $\mu$eV. It was found that, due to proximity to a quantum critical point among the FM, AFM, and spin liquid phases ~\cite{Yan2017}, $\mathcal{S}(\mathbf{q}, \omega, T)$ exhibits a dynamical scaling behavior 
\begin{equation}\label{eq:scaling}
    k_B T \mathcal{S}(\mathbf{q}, \omega, T) \sim 2 \left( \frac{1}{\exp (x) - 1} +1  \right) \frac{B x}{R^2 + x^2},
\end{equation}
where $x = \hbar \omega/k_B T$ with fitted parameters $B = 0.0181$ and $R = 0.80$ \cite{scheie2022dynamical}. By inserting this scaling form into Eq.~(\ref{eq:tau}), one can calculate the reversed relaxation time of the charge carriers due to the YbTO QSF. After performing the frequency integral up to a small cutoff at $4.3 ~{\mu}$eV, we indeed arrived at a $1/T$-like behavior of the reversed relaxation time as shown in Fig. \ref{fig2}(c), resembling the $1/T$ behavior of the BIO resistivity. This behavior is robust against different cutoff values in a wide range, as demonstrated in Fig. S6 \cite{supplementary} with the cutoff from $0.086 ~{\mu}$eV to $8.6 ~{\mu}$eV.
The temperature dependence of the BIO resistivity therefore reflects the scaling behavior of the QSF in YbTO. 

\begin{figure*}[t]
    \centering
    \includegraphics[width=1\textwidth]{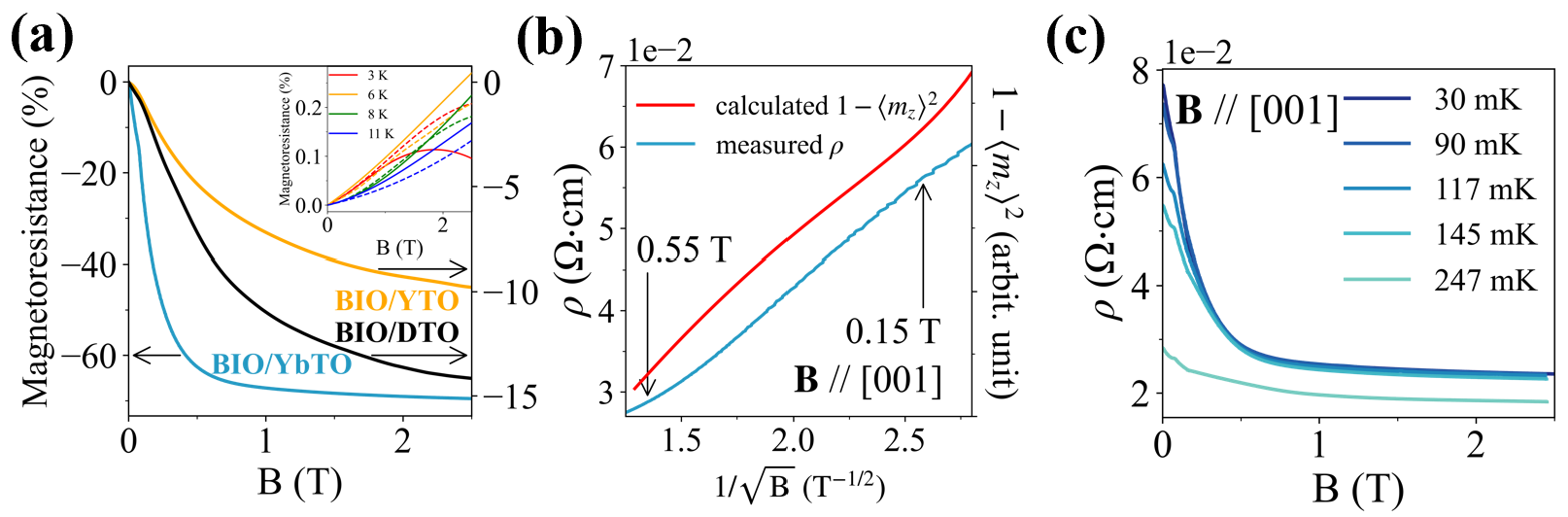}
    \caption{Field-dependent resistivity of BIO/YbTO. (a) MR of BIO/YbTO, BIO/DTO and BIO/YTO at 30 mK with B // [001]. Inset: MR of BIO/YbTO (solid lines) and BIO/YTO (dashed lines) above 3 K with B // [001]. (b) Blue line: The resistivity of BIO/YbTO at 30 mK with B // [001] plotted against $\mathrm{1/\sqrt{B}}$ for the low field region where the MR shows the strongest negative drop. Red line: The calculated $1 - \langle m_z \rangle ^2$ that represents the amplitude of the transverse spin fluctuations shows a similar $\mathrm{1/\sqrt{B}}$ dependence as the resistivity does. (c) Field dependence of the resistivity of BIO/YbTO with B // [001] at four temperatures below $T_{c}$ of YbTO and one slightly above.} 
    \label{fig3}
\end{figure*}
This proximitized transport behavior can be tuned by magnetic field. When applying a field along [001] to suppress the YbTO QSF at 30 mK, the BIO resistivity decreases rapidly below $1$~T [Fig. \ref{fig3}(a)]. The decrease quickly slows down as the field approaches 1 T, resulting in a nearly flat magnetoresistance (MR) curve beyond that point. This rapid recovery of the conductivity can be understood from the fact that the AFM fluctuations are strongly suppressed by the external magnetic field through enforcing the FM order \cite{Scheie2020a}. In fact, the broad spin waves of YbTO have been shown to disappear quickly under magnetic field by inelastic neutron scattering \cite{Scheie2020a}. This sharp negative MR by tens of percents at small fields is also in remarkable contrast with the gradual and slow decrease observed in the BIO/YTO reference sample across the entire measured field range. This distinction disappears at elevated temperatures above 3 K, when thermal fluctuations dominate over quantum fluctuations, and the MR of BIO/YTO and BIO/YbTO becomes virtually the same as seen in the inset of [Fig. \ref{fig3}(a)]. One can make a similar comparison with the reported BIO/DTO heterostructure\cite{zhang2023anomalous}. They both have a positive sign and a similar amplitude around 0.2\% at high temperatures. Such a similarity again points to the dominant role of the thermal fluctuations in the BIO transport properties at high temperatures. Their difference emerges at low temperatures where thermal fluctuations are suppressed. 
At 30 mK, other than a relatively small anomaly ($\sim1\%$) when the field induces the ice rule-breaking transition of classical spin ice in DTO, the MR of BIO/DTO is overall very similar to BIO/YTO and thus distinct from the large sharp negative MR observed here for BIO/YbTO [Fig. \ref{fig3}(a)]. This comparison highlights the drastically enhanced impact of quantum spins on the proximitized transport.

Quantitatively, we find a $\mathrm{1/\sqrt{B}}$ scaling for the sharp resistivity drop [Fig. \ref{fig3}(b)] through logarithmic fitting of the MR at 30 mK, which is also absent in BIO/YTO [Fig. S7(a)] \cite{supplementary}. 
Again, such an algebraic scaling cannot be explained by conventional spin waves, as the Bose-Einstein distribution would result in exponentially suppression of spin fluctuations due to the low temperature (30 mK) and the field-induced gap \cite{fuchs2005spin}. Instead, we resort to a simple model that has been used to capture the magnetic-field dependence of the spin fluctuations of YbTO at the mean-field level~\cite{saubert2020}. Specifically, taking a coarse-grained magnetization parameterized by a unit vector $\mathbf{m}$ leads to the following potential energy that describes the competition between the Zeeman energy and the magnetic anisotropy~\cite{saubert2020}:
\begin{equation}
\label{eq:cubic-anisotropy}
    U = - \mathbf{B} \cdot \mathbf{m} -K_1 \sum_i (\mathbf{m} \cdot \mathbf{e}_i)^4 
     - K_2 \prod_i (\mathbf{m} \cdot \mathbf{e}_i)^2.
\end{equation}
The form of the anisotropy respects the cubic symmetry of YbTO, and the parameters $K_1 = 0.14$ and $K_2 = -0.55$ have been used to match the magnetic-field induced phase transitions and the corresponding critical field strengths \cite{saubert2020}. 
For a magnetic field applied along the $z$ direction (i.e., [001]), the thermal average $1-\langle m_z \rangle^2$ measures the magnitude of the allowed transverse fluctuations. One can see in Fig. \ref{fig3}(b) that it increases with decreasing field and roughly reproduces the $\mathrm{1/\sqrt{B}}$ scaling observed in MR.
The effective cubic anisotropy here can be viewed as a coarse-grained result of interactions between the four spins on a single Yb tetrahedron,
suggesting that short-range correlations largely contribute to the fluctuations and the proximitized MR. 

The combined temperature-magnetic field control on the proximitized transport is presented in Fig. \ref{fig3}(c), where the unnormalized MR curves at different temperatures all flatten above 1 T and clearly converge to a similar level. In other words, the strong temperature dependence of the resistivity only occurs around zero field, and the temperature dependence becomes minimal once the external magnetic field effectively suppresses spin fluctuations. As a result, the negative MR at small fields is most pronounced at lower temperatures, while thermal fluctuations reduce its sensitivity. These findings provide compelling evidence that the BIO resistivity offers an effective means of assessing the magnitudes of YbTO QSF.

\subsection{Proximitized Transport due to Field-Induced Transition}

\begin{figure*}[t]
    \centering
    \includegraphics[width=0.8\textwidth]{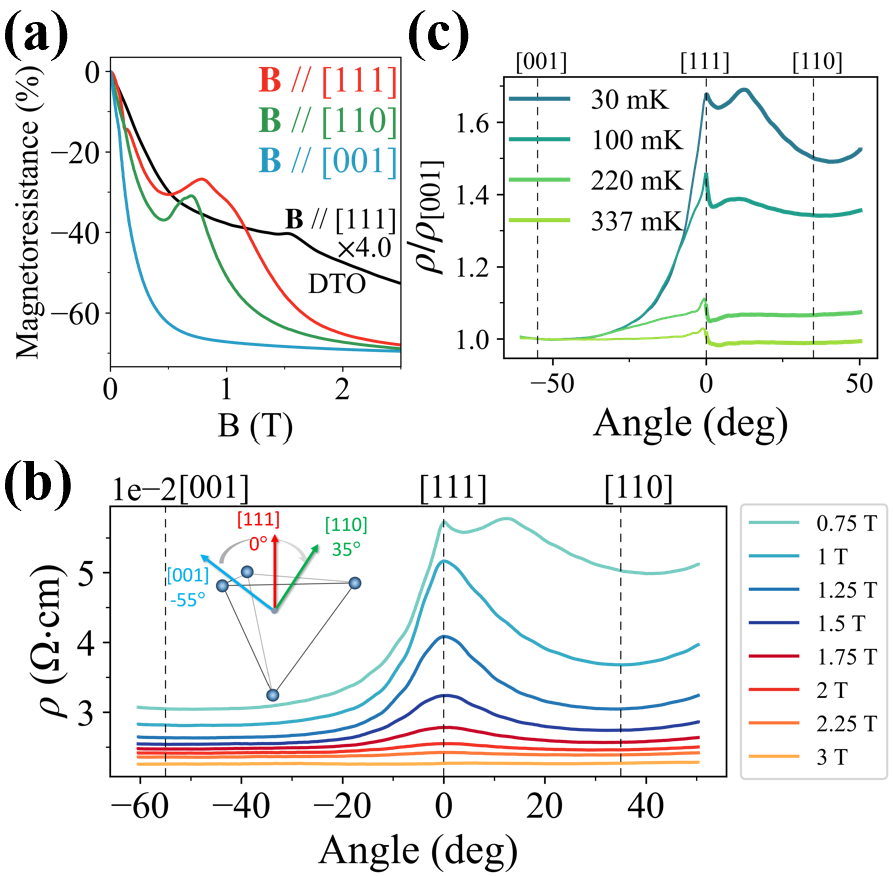}
    \caption{Angle-dependent MR of BIO/YbTO. (a) MR at 30 mK with magnetic field applied along three high-symmetric directions. Black line is MR of BIO/DTO at 30 mK with magnetic field applied along [111] axis. This black line is multiplied with a factor of 4 for better visualization. (b) Field angular scans of the resistivity at 30 mK under seven different field strengths above the critical field of B // [111] and one field strength slightly below. Insert: The measurement configuration. The applied field is vertical and the current is along [1-10]. (c) Field angular scans of the resistivity at 0.75 T at three temperatures below $T_{c}$ and one above. For the purpose of comparison, the scans are normalized by the resistivity with B // [001] at each temperature.}\label{fig4}
\end{figure*}

The proximitized transport strongly responds to the field-induced transitions of YbTO as well. Note that both BIO film and crystal show the linear isotropic MR above 2 K \cite{chu2019}. Significant anisotropy arises here for BIO/YbTO. With B // [110] and B // [111] [Fig. \ref{fig4}(a)], the MR initially shows a sharp negative response similar to B // [001]. However, the rate of decrease slows down followed by an upturn. As the field continues to increase, the MR becomes negative again, resulting in a broad peak-like feature. 
These anisotropic behaviors are absent in the BIO/YTO reference sample [Fig. S7(b)] \cite{supplementary}, indicating that they originate from the YbTO quantum magnetism. The locations of the peak-like feature around 0.7 T and 0.79 T for B // [110] and B // [111], respectively, well align with the critical fields of the YbTO phase transition \cite{scheie2017reentrant,saubert2020} driven by the competition between the cubic magnetic anisotropy and the Zeeman energy \cite{ross2011} described above by Eq.~(\ref{eq:cubic-anisotropy}). Specifically, since $\textlangle$001$\textrangle$ is the easy axis of the canted FM order, a [111] field is an effective transverse field that enforces a polarized state by mixing the three spontaneous $\textlangle$001$\textrangle$ configurations, causing enhanced fluctuations which are now captured by the BIO resistivity. For comparison, this peak-like feature is much more pronounced than the MR anomaly due to the classical spin-ice transition in the BIO/DTO heterostructure \cite{zhang2023anomalous}, showcasing again the much stronger impacts of quantum spin systems than classical spin systems on proximitized transport.  

The pronounced peak-like feature must be driven by is the YbTO QSFs rather than its stray field. The stray field near the boundary of a magnetic material is proportional to the magnetization. The magnetization of DTO along $\textlangle$111$\textrangle$ (both the Kagome ice state and the 3-in-1-out state) is much larger than the magnetization of YbTO of all directions \cite{saubert2020, fukazawa2002magnetic}. Therefore, if the stray field dictates the size of the anomaly or any feature in MR, one would expect it to be much more significant in BIO/DTO than BIO/YbTO. Moreover, if an MR feature is driven by the stray field, it should resemble the shape of the M(B) curve, which means the BIO/DTO anomaly would have a step-like shape because of the sharp magnetization jump. Meanwhile, the M(B) curve of YbTO has no jump but just changes slope via a kink at the critical field. This expectation based on stray field is clearly inconsistent with the experimental observations of the MR measurements.

At field above 2 T, both MR curves with B // [110] and B // [111] become flatten deep into the field-polarized state where the fluctuations are completely suppressed. They also converge to the same level as B // [001]. Such weak dependence on the field direction at high fields shows that the orientation of the magnetization of YbTO has little impact on charge transport in BIO despite the magnetic anisotropy, which can be explained by the fact that the field-polarized state can be described as a classical order parameter. This observation was further confirmed by angular scans that continuously rotate the field from [001] to [111], and to [110]. As seen in [Fig. \ref{fig4}(b)], 
the resistivity displayed almost no angle dependence at 3 T. Only when the field decreases toward the critical field, the angle dependence becomes dramatically stronger, with the resistivity showing a rapid increase near [111] where the fluctuations are the strongest. The resistance changed smoothly with the field angle without any kink, indicating that the field is continuously tuning the strength of spin fluctuations. 

What is also particularly interesting is that when the field dropped below the critical field of [111], as shown at 0.75 T, a sharp dip in the angular change around [111] was observed, leading to a double-peak-like shape. This behavior can be attributed to the domain conversion of the spontaneous canted FM order since [111] is the hard axis and thus the tipping point of domain re-population \cite{scheie2017reentrant}. The dip likely marks the narrow angular range of domain coexistence. More importantly, as seen in Fig. \ref{fig4}(c), this dip persists at temperatures above $T_{c}$, capturing the famous reentrant behavior of the FM state in the [111] field-temperature phase diagram \cite{scheie2017reentrant} that is a signature of the strong QSF in YbTO. All the transport behaviors discussed above are well reproducible in different samples (Fig. S8 \cite{supplementary}).

\subsection{Impacts of YbTO Nonstoichiometry}
\begin{figure}[t]
    \centering
    \centering{\includegraphics[clip,width=8.6cm]{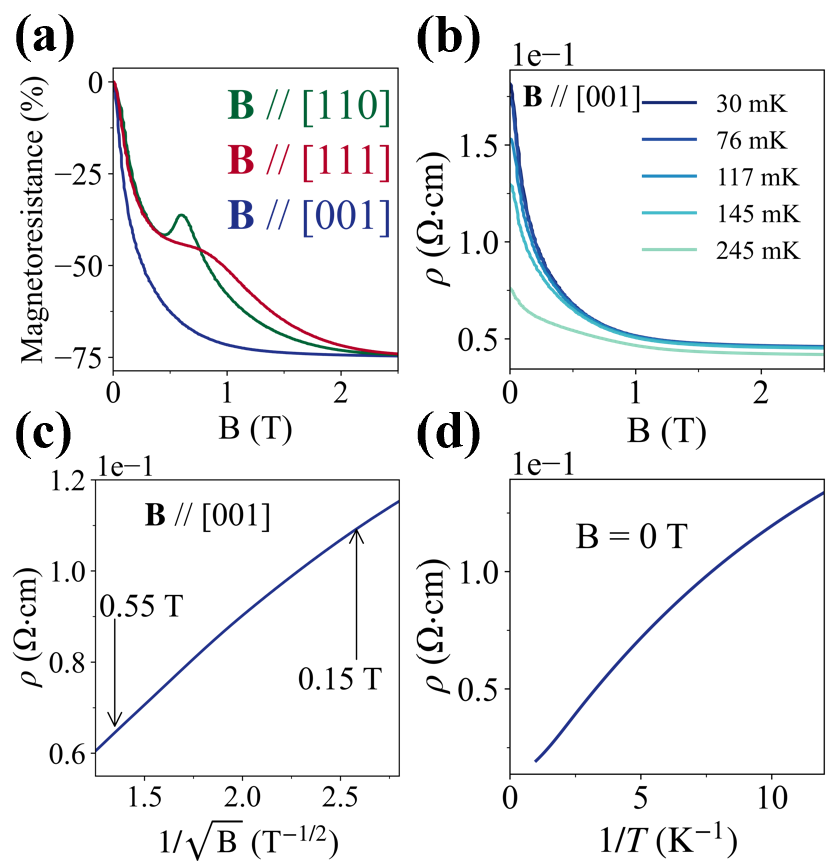}}
    \caption{Resistivity behavior of a 4.2 nm thick BIO film on YbTO-ns. (a) MR at 30 mK with magnetic field applied along three high-symmetric directions shows the initial sharp negative drop. It flattens with increasing field when B // [001], whereas the additional features appear at the critical fields for B // [110] and B // [111]. (b) Field dependence of the resistivity with B // [001] at five temperatures similar to BIO/YbTO shown in Fig.~\ref{fig3}(c). (c) $\mathrm{1/\sqrt{B}}$ dependence of the resistivity at 30 mK with B // [001]. (d) $1/T$ dependence of the resistivity at zero field. Compared with BIO/YbTO [Fig.~\ref{fig2}(b)], a sublinear deviation can be seen.
    }
    \label{fig5}
\end{figure}

Given the remarkable sensitivity of the proximitized transport to the YbTO quantum magnetism, we performed the same measurements on a BIO film grown on a nonstoichiometric YbTO (YbTO-ns) substrate (See Supplementary \cite{supplementary}) to investigate the impact of site disorder between the \ch{Yb^{3+}} and \ch{Ti^{4+}} ions. This disorder is known to effectively suppress the spontaneous FM order and further broaden the spin wave spectra \cite{Arpino2017, Scheie2020a}. Fig. \ref{fig5}(a-c) shows that the MR is overall similar to that in the stoichiometric case, including the sharp negative MR, the $\rm{1/\sqrt{B}}$ dependence, and the response to the field-induced transition. These results are consistent with the resistive contribution from short-range fluctuations. Notably, the unnormalized MR with B // [001] reveals an enhanced resistivity at zero-field and low temperatures, indicating the presence of significant QSF despite the disorder. However, a sub-linear deviation from the $1/T$ scaling can be observed in Fig. \ref{fig5}(d), suggesting that the temperature dependence of the resistivity is more sensitive to the spatial coherence in the spin fluctuations, which may arise from either the inhomogeneity of the exchange interactions or an overall deviation from the quantum critical regime. In fact, the dynamical scaling behavior of YbTO described by Eq.~(\ref{eq:scaling}) was observed in stoichiometric YbTO crystals \cite{scheie2022dynamical}.

\section*{Discussion}
It has been a long-standing puzzle at what levels coherent quantum fluctuations play a role in materials close to realizing exotic quantum spin liquids. In certain candidate systems, the extremely broad excitation spectrum may be understood as spin waves from magnetically disordered ground states~\cite{zhang2019}, and spin fluctuations mainly consist of thermal occupation of spin-wave modes. Recent inspirations are offered along the lines of witnessing quantum entanglement through magnetic spectrum~\cite{Laurell2021}, anomalous thermal transport experiments~\cite{hirschberger2019enhanced}, and magnetic noise characterization~\cite{chatterjee2019}, etc. 
Proximitized electric transport study of QSF demonstrated in this work offers several advantages, such as a higher temperature resolution and an enhanced sensitivity to low-energy fluctuations with large-momentum transfer, where scattering processes are particularly efficient in reducing the electron relaxation time.

It is remarkable to see that the thin film resistivity of our heterostructure construction well captures the essential properties of the bulk quantum magnetism, despite the potential complications from the interface. 
The same transition temperature and critical field between proximitized transport and susceptibility indicate the Yb-Yb interface magnetism is at least qualitatively the same as the bulk. Note that all the matrix elements of the Yb-Yb exchange matrix are finite for bulk YbTO \cite{ross2011}, and they account for all symmetry components of the exchange interaction, i.e., XY-like ($J_1$), Ising-like ($J_2$), pseudodipolar ($J_3$), and DM ($J_4$) interactions. Any quantitative difference at the interface could be accounted for by modifications of these parameters. The experimental facts discussed above indicate that such changes at the interface (if any) make no significant difference, at least from the prospect of the proximitized transport.

In summary, our investigation on the electric transport properties of a BIO film deposited on a YbTO single crystal shows a large enhancement in resistance due to the proximity to quantum magnetism at low temperatures. The observed scaling behaviors of the resistance with temperature and magnetic field strongly point to the mechanism of QSF-mediated scattering of electrons at the interface. Moreover, the distinct responses under different field directions highlights how the MR of BIO accurately captures the magnetic anisotropy of the field-induced quantum phase transitions in YbTO. Overall, our study establishes a novel approach to realize strong correlation between the electronic transport and the QSF by employing a heterostructure comprising a metallic film epitaxially grown on a quantum magnet, which provides guidelines for future exploration of proximitized transport phenomena derived from insulting quantum magnets. Furthermore, the thermal and magnetic controls of the proximitized transport hold promise for potential applications in spintronics by leveraging the unique properties of quantum magnetism.

\section*{ACKNOWLEDGMENTS}
This research is supported by the U.S. Department of Energy under grant No. DE-SC0020254. The authors thank Allen Scheie and Cristian Batista for fruitful discussions. The authors thank Jenia Karapetrova for assistance with the synchrotron X-ray diffraction and thank Zhigang Jiang for assistance with the atomic-force-microscope measurement. C. X. acknowledges support from the Center for Material Processing at the University of Tennessee, Knoxville. The synchrotron X-ray diffraction measurement used resources of the Advanced Photon Source, a U.S. Department of Energy (DOE) Office of Science User Facility operated for the DOE Office of Science by Argonne National Laboratory under Contract No. DE-AC02-06CH11357. A portion of this work was performed at the National High Magnetic Field Laboratory, which is supported by the National Science Foundation Cooperative Agreement No. DMR-1644779 and the state of Florida. Y. Z. is supported by the start-up fund at the University of Tennessee.

\nocite{li2013single}
\nocite{zhang2023anomalous}
\nocite{yang2017epitaxial}

	\nocite{apsrev42Control}

	\bibliography{reference}

\begin{thebibliography}{63}%
\makeatletter
\providecommand \@ifxundefined [1]{%
 \@ifx{#1\undefined}
}%
\providecommand \@ifnum [1]{%
 \ifnum #1\expandafter \@firstoftwo
 \else \expandafter \@secondoftwo
 \fi
}%
\providecommand \@ifx [1]{%
 \ifx #1\expandafter \@firstoftwo
 \else \expandafter \@secondoftwo
 \fi
}%
\providecommand \natexlab [1]{#1}%
\providecommand \enquote  [1]{``#1''}%
\providecommand \bibnamefont  [1]{#1}%
\providecommand \bibfnamefont [1]{#1}%
\providecommand \citenamefont [1]{#1}%
\providecommand \href@noop [0]{\@secondoftwo}%
\providecommand \href [0]{\begingroup \@sanitize@url \@href}%
\providecommand \@href[1]{\@@startlink{#1}\@@href}%
\providecommand \@@href[1]{\endgroup#1\@@endlink}%
\providecommand \@sanitize@url [0]{\catcode `\\12\catcode `\$12\catcode `\&12\catcode `\#12\catcode `\^12\catcode `\_12\catcode `\%12\relax}%
\providecommand \@@startlink[1]{}%
\providecommand \@@endlink[0]{}%
\providecommand \url  [0]{\begingroup\@sanitize@url \@url }%
\providecommand \@url [1]{\endgroup\@href {#1}{\urlprefix }}%
\providecommand \urlprefix  [0]{URL }%
\providecommand \Eprint [0]{\href }%
\providecommand \doibase [0]{https://doi.org/}%
\providecommand \selectlanguage [0]{\@gobble}%
\providecommand \bibinfo  [0]{\@secondoftwo}%
\providecommand \bibfield  [0]{\@secondoftwo}%
\providecommand \translation [1]{[#1]}%
\providecommand \BibitemOpen [0]{}%
\providecommand \bibitemStop [0]{}%
\providecommand \bibitemNoStop [0]{.\EOS\space}%
\providecommand \EOS [0]{\spacefactor3000\relax}%
\providecommand \BibitemShut  [1]{\csname bibitem#1\endcsname}%
\let\auto@bib@innerbib\@empty
\bibitem [{\citenamefont {Wolf}\ \emph {et~al.}(2001)\citenamefont {Wolf}, \citenamefont {Awschalom}, \citenamefont {Buhrman}, \citenamefont {Daughton}, \citenamefont {von Moln{\'a}r}, \citenamefont {Roukes}, \citenamefont {Chtchelkanova},\ and\ \citenamefont {Treger}}]{wolf2001spintronics}%
  \BibitemOpen
  \bibfield  {author} {\bibinfo {author} {\bibfnamefont {S.}~\bibnamefont {Wolf}}, \bibinfo {author} {\bibfnamefont {D.}~\bibnamefont {Awschalom}}, \bibinfo {author} {\bibfnamefont {R.}~\bibnamefont {Buhrman}}, \bibinfo {author} {\bibfnamefont {J.}~\bibnamefont {Daughton}}, \bibinfo {author} {\bibfnamefont {v.~S.}\ \bibnamefont {von Moln{\'a}r}}, \bibinfo {author} {\bibfnamefont {M.}~\bibnamefont {Roukes}}, \bibinfo {author} {\bibfnamefont {A.~Y.}\ \bibnamefont {Chtchelkanova}},\ and\ \bibinfo {author} {\bibfnamefont {D.}~\bibnamefont {Treger}},\ }\bibfield  {title} {\bibinfo {title} {Spintronics: a spin-based electronics vision for the future},\ }\href@noop {} {\bibfield  {journal} {\bibinfo  {journal} {Science}\ }\textbf {\bibinfo {volume} {294}},\ \bibinfo {pages} {1488} (\bibinfo {year} {2001})}\BibitemShut {NoStop}%
\bibitem [{\citenamefont {Hirohata}\ and\ \citenamefont {Takanashi}(2014)}]{hirohata2014magnetic}%
  \BibitemOpen
  \bibfield  {author} {\bibinfo {author} {\bibfnamefont {A.}~\bibnamefont {Hirohata}}\ and\ \bibinfo {author} {\bibfnamefont {K.}~\bibnamefont {Takanashi}},\ }\bibfield  {title} {\bibinfo {title} {Future perspectives for spintronic devices},\ }\href@noop {} {\bibfield  {journal} {\bibinfo  {journal} {Journal of Physics D: Applied Physics}\ }\textbf {\bibinfo {volume} {47}},\ \bibinfo {pages} {193001} (\bibinfo {year} {2014})}\BibitemShut {NoStop}%
\bibitem [{\citenamefont {Baltz}\ \emph {et~al.}(2018)\citenamefont {Baltz}, \citenamefont {Manchon}, \citenamefont {Tsoi}, \citenamefont {Moriyama}, \citenamefont {Ono},\ and\ \citenamefont {Tserkovnyak}}]{baltz2018antiferromagnetic}%
  \BibitemOpen
  \bibfield  {author} {\bibinfo {author} {\bibfnamefont {V.}~\bibnamefont {Baltz}}, \bibinfo {author} {\bibfnamefont {A.}~\bibnamefont {Manchon}}, \bibinfo {author} {\bibfnamefont {M.}~\bibnamefont {Tsoi}}, \bibinfo {author} {\bibfnamefont {T.}~\bibnamefont {Moriyama}}, \bibinfo {author} {\bibfnamefont {T.}~\bibnamefont {Ono}},\ and\ \bibinfo {author} {\bibfnamefont {Y.}~\bibnamefont {Tserkovnyak}},\ }\bibfield  {title} {\bibinfo {title} {Antiferromagnetic spintronics},\ }\href@noop {} {\bibfield  {journal} {\bibinfo  {journal} {Reviews of Modern Physics}\ }\textbf {\bibinfo {volume} {90}},\ \bibinfo {pages} {015005} (\bibinfo {year} {2018})}\BibitemShut {NoStop}%
\bibitem [{\citenamefont {Jungwirth}\ \emph {et~al.}(2016)\citenamefont {Jungwirth}, \citenamefont {Marti}, \citenamefont {Wadley},\ and\ \citenamefont {Wunderlich}}]{jungwirth2016antiferromagnetic}%
  \BibitemOpen
  \bibfield  {author} {\bibinfo {author} {\bibfnamefont {T.}~\bibnamefont {Jungwirth}}, \bibinfo {author} {\bibfnamefont {X.}~\bibnamefont {Marti}}, \bibinfo {author} {\bibfnamefont {P.}~\bibnamefont {Wadley}},\ and\ \bibinfo {author} {\bibfnamefont {J.}~\bibnamefont {Wunderlich}},\ }\bibfield  {title} {\bibinfo {title} {Antiferromagnetic spintronics},\ }\href@noop {} {\bibfield  {journal} {\bibinfo  {journal} {Nature nanotechnology}\ }\textbf {\bibinfo {volume} {11}},\ \bibinfo {pages} {231} (\bibinfo {year} {2016})}\BibitemShut {NoStop}%
\bibitem [{\citenamefont {Gomonay}\ and\ \citenamefont {Loktev}(2014)}]{gomonay2014spintronics}%
  \BibitemOpen
  \bibfield  {author} {\bibinfo {author} {\bibfnamefont {E.}~\bibnamefont {Gomonay}}\ and\ \bibinfo {author} {\bibfnamefont {V.}~\bibnamefont {Loktev}},\ }\bibfield  {title} {\bibinfo {title} {Spintronics of antiferromagnetic systems},\ }\href@noop {} {\bibfield  {journal} {\bibinfo  {journal} {Low Temperature Physics}\ }\textbf {\bibinfo {volume} {40}},\ \bibinfo {pages} {17} (\bibinfo {year} {2014})}\BibitemShut {NoStop}%
\bibitem [{\citenamefont {Sachdev}(1999)}]{sachdev1999quantum}%
  \BibitemOpen
  \bibfield  {author} {\bibinfo {author} {\bibfnamefont {S.}~\bibnamefont {Sachdev}},\ }\bibfield  {title} {\bibinfo {title} {Quantum phase transitions},\ }\href@noop {} {\bibfield  {journal} {\bibinfo  {journal} {Physics world}\ }\textbf {\bibinfo {volume} {12}},\ \bibinfo {pages} {33} (\bibinfo {year} {1999})}\BibitemShut {NoStop}%
\bibitem [{\citenamefont {Balents}(2010)}]{balents2010spin}%
  \BibitemOpen
  \bibfield  {author} {\bibinfo {author} {\bibfnamefont {L.}~\bibnamefont {Balents}},\ }\bibfield  {title} {\bibinfo {title} {Spin liquids in frustrated magnets},\ }\href@noop {} {\bibfield  {journal} {\bibinfo  {journal} {Nature}\ }\textbf {\bibinfo {volume} {464}},\ \bibinfo {pages} {199} (\bibinfo {year} {2010})}\BibitemShut {NoStop}%
\bibitem [{\citenamefont {Zhou}\ \emph {et~al.}(2017)\citenamefont {Zhou}, \citenamefont {Kanoda},\ and\ \citenamefont {Ng}}]{zhou2017quantum}%
  \BibitemOpen
  \bibfield  {author} {\bibinfo {author} {\bibfnamefont {Y.}~\bibnamefont {Zhou}}, \bibinfo {author} {\bibfnamefont {K.}~\bibnamefont {Kanoda}},\ and\ \bibinfo {author} {\bibfnamefont {T.-K.}\ \bibnamefont {Ng}},\ }\bibfield  {title} {\bibinfo {title} {Quantum spin liquid states},\ }\href@noop {} {\bibfield  {journal} {\bibinfo  {journal} {Reviews of Modern Physics}\ }\textbf {\bibinfo {volume} {89}},\ \bibinfo {pages} {025003} (\bibinfo {year} {2017})}\BibitemShut {NoStop}%
\bibitem [{\citenamefont {Broholm}\ \emph {et~al.}(2020)\citenamefont {Broholm}, \citenamefont {Cava}, \citenamefont {Kivelson}, \citenamefont {Nocera}, \citenamefont {Norman},\ and\ \citenamefont {Senthil}}]{broholm2020quantum}%
  \BibitemOpen
  \bibfield  {author} {\bibinfo {author} {\bibfnamefont {C.}~\bibnamefont {Broholm}}, \bibinfo {author} {\bibfnamefont {R.}~\bibnamefont {Cava}}, \bibinfo {author} {\bibfnamefont {S.}~\bibnamefont {Kivelson}}, \bibinfo {author} {\bibfnamefont {D.}~\bibnamefont {Nocera}}, \bibinfo {author} {\bibfnamefont {M.}~\bibnamefont {Norman}},\ and\ \bibinfo {author} {\bibfnamefont {T.}~\bibnamefont {Senthil}},\ }\bibfield  {title} {\bibinfo {title} {Quantum spin liquids},\ }\href@noop {} {\bibfield  {journal} {\bibinfo  {journal} {Science}\ }\textbf {\bibinfo {volume} {367}},\ \bibinfo {pages} {0668} (\bibinfo {year} {2020})}\BibitemShut {NoStop}%
\bibitem [{\citenamefont {Stone}\ \emph {et~al.}(2006)\citenamefont {Stone}, \citenamefont {Zaliznyak}, \citenamefont {Hong}, \citenamefont {Broholm},\ and\ \citenamefont {Reich}}]{stone2006quasiparticle}%
  \BibitemOpen
  \bibfield  {author} {\bibinfo {author} {\bibfnamefont {M.~B.}\ \bibnamefont {Stone}}, \bibinfo {author} {\bibfnamefont {I.~A.}\ \bibnamefont {Zaliznyak}}, \bibinfo {author} {\bibfnamefont {T.}~\bibnamefont {Hong}}, \bibinfo {author} {\bibfnamefont {C.~L.}\ \bibnamefont {Broholm}},\ and\ \bibinfo {author} {\bibfnamefont {D.~H.}\ \bibnamefont {Reich}},\ }\bibfield  {title} {\bibinfo {title} {Quasiparticle breakdown in a quantum spin liquid},\ }\href@noop {} {\bibfield  {journal} {\bibinfo  {journal} {Nature}\ }\textbf {\bibinfo {volume} {440}},\ \bibinfo {pages} {187} (\bibinfo {year} {2006})}\BibitemShut {NoStop}%
\bibitem [{\citenamefont {Nayak}\ \emph {et~al.}(2008)\citenamefont {Nayak}, \citenamefont {Simon}, \citenamefont {Stern}, \citenamefont {Freedman},\ and\ \citenamefont {Sarma}}]{nayak2008non}%
  \BibitemOpen
  \bibfield  {author} {\bibinfo {author} {\bibfnamefont {C.}~\bibnamefont {Nayak}}, \bibinfo {author} {\bibfnamefont {S.~H.}\ \bibnamefont {Simon}}, \bibinfo {author} {\bibfnamefont {A.}~\bibnamefont {Stern}}, \bibinfo {author} {\bibfnamefont {M.}~\bibnamefont {Freedman}},\ and\ \bibinfo {author} {\bibfnamefont {S.~D.}\ \bibnamefont {Sarma}},\ }\bibfield  {title} {\bibinfo {title} {Non-abelian anyons and topological quantum computation},\ }\href@noop {} {\bibfield  {journal} {\bibinfo  {journal} {Reviews of Modern Physics}\ }\textbf {\bibinfo {volume} {80}},\ \bibinfo {pages} {1083} (\bibinfo {year} {2008})}\BibitemShut {NoStop}%
\bibitem [{\citenamefont {Savary}\ and\ \citenamefont {Balents}(2016)}]{savary2016quantum}%
  \BibitemOpen
  \bibfield  {author} {\bibinfo {author} {\bibfnamefont {L.}~\bibnamefont {Savary}}\ and\ \bibinfo {author} {\bibfnamefont {L.}~\bibnamefont {Balents}},\ }\bibfield  {title} {\bibinfo {title} {Quantum spin liquids: a review},\ }\href@noop {} {\bibfield  {journal} {\bibinfo  {journal} {Reports on Progress in Physics}\ }\textbf {\bibinfo {volume} {80}},\ \bibinfo {pages} {016502} (\bibinfo {year} {2016})}\BibitemShut {NoStop}%
\bibitem [{\citenamefont {Taillefumier}\ \emph {et~al.}(2017)\citenamefont {Taillefumier}, \citenamefont {Benton}, \citenamefont {Yan}, \citenamefont {Jaubert},\ and\ \citenamefont {Shannon}}]{taillefumier2017competing}%
  \BibitemOpen
  \bibfield  {author} {\bibinfo {author} {\bibfnamefont {M.}~\bibnamefont {Taillefumier}}, \bibinfo {author} {\bibfnamefont {O.}~\bibnamefont {Benton}}, \bibinfo {author} {\bibfnamefont {H.}~\bibnamefont {Yan}}, \bibinfo {author} {\bibfnamefont {L.~D.}\ \bibnamefont {Jaubert}},\ and\ \bibinfo {author} {\bibfnamefont {N.}~\bibnamefont {Shannon}},\ }\bibfield  {title} {\bibinfo {title} {Competing spin liquids and hidden spin-nematic order in spin ice with frustrated transverse exchange},\ }\href@noop {} {\bibfield  {journal} {\bibinfo  {journal} {Physical Review X}\ }\textbf {\bibinfo {volume} {7}},\ \bibinfo {pages} {041057} (\bibinfo {year} {2017})}\BibitemShut {NoStop}%
\bibitem [{\citenamefont {Takahashi}\ \emph {et~al.}(2019)\citenamefont {Takahashi}, \citenamefont {Wang}, \citenamefont {Arsenault}, \citenamefont {Imai}, \citenamefont {Abramchuk}, \citenamefont {Tafti},\ and\ \citenamefont {Singer}}]{takahashi2019spin}%
  \BibitemOpen
  \bibfield  {author} {\bibinfo {author} {\bibfnamefont {S.~K.}\ \bibnamefont {Takahashi}}, \bibinfo {author} {\bibfnamefont {J.}~\bibnamefont {Wang}}, \bibinfo {author} {\bibfnamefont {A.}~\bibnamefont {Arsenault}}, \bibinfo {author} {\bibfnamefont {T.}~\bibnamefont {Imai}}, \bibinfo {author} {\bibfnamefont {M.}~\bibnamefont {Abramchuk}}, \bibinfo {author} {\bibfnamefont {F.}~\bibnamefont {Tafti}},\ and\ \bibinfo {author} {\bibfnamefont {P.~M.}\ \bibnamefont {Singer}},\ }\bibfield  {title} {\bibinfo {title} {Spin excitations of a proximate kitaev quantum spin liquid realized in $\ch{Cu2IrO3}$},\ }\href@noop {} {\bibfield  {journal} {\bibinfo  {journal} {Physical Review X}\ }\textbf {\bibinfo {volume} {9}},\ \bibinfo {pages} {031047} (\bibinfo {year} {2019})}\BibitemShut {NoStop}%
\bibitem [{\citenamefont {Aasen}\ \emph {et~al.}(2020)\citenamefont {Aasen}, \citenamefont {Mong}, \citenamefont {Hunt}, \citenamefont {Mandrus},\ and\ \citenamefont {Alicea}}]{aasen2020electrical}%
  \BibitemOpen
  \bibfield  {author} {\bibinfo {author} {\bibfnamefont {D.}~\bibnamefont {Aasen}}, \bibinfo {author} {\bibfnamefont {R.~S.}\ \bibnamefont {Mong}}, \bibinfo {author} {\bibfnamefont {B.~M.}\ \bibnamefont {Hunt}}, \bibinfo {author} {\bibfnamefont {D.}~\bibnamefont {Mandrus}},\ and\ \bibinfo {author} {\bibfnamefont {J.}~\bibnamefont {Alicea}},\ }\bibfield  {title} {\bibinfo {title} {Electrical probes of the non-abelian spin liquid in kitaev materials},\ }\href@noop {} {\bibfield  {journal} {\bibinfo  {journal} {Physical Review X}\ }\textbf {\bibinfo {volume} {10}},\ \bibinfo {pages} {031014} (\bibinfo {year} {2020})}\BibitemShut {NoStop}%
\bibitem [{\citenamefont {Kitaev}(2006)}]{kitaev2006anyons}%
  \BibitemOpen
  \bibfield  {author} {\bibinfo {author} {\bibfnamefont {A.}~\bibnamefont {Kitaev}},\ }\bibfield  {title} {\bibinfo {title} {Anyons in an exactly solved model and beyond},\ }\href@noop {} {\bibfield  {journal} {\bibinfo  {journal} {Annals of Physics}\ }\textbf {\bibinfo {volume} {321}},\ \bibinfo {pages} {2} (\bibinfo {year} {2006})}\BibitemShut {NoStop}%
\bibitem [{\citenamefont {Banerjee}\ \emph {et~al.}(2016)\citenamefont {Banerjee}, \citenamefont {Bridges}, \citenamefont {Yan}, \citenamefont {Aczel}, \citenamefont {Li}, \citenamefont {Stone}, \citenamefont {Granroth}, \citenamefont {Lumsden}, \citenamefont {Yiu}, \citenamefont {Knolle} \emph {et~al.}}]{banerjee2017proximate}%
  \BibitemOpen
  \bibfield  {author} {\bibinfo {author} {\bibfnamefont {A.}~\bibnamefont {Banerjee}}, \bibinfo {author} {\bibfnamefont {C.}~\bibnamefont {Bridges}}, \bibinfo {author} {\bibfnamefont {J.-Q.}\ \bibnamefont {Yan}}, \bibinfo {author} {\bibfnamefont {A.}~\bibnamefont {Aczel}}, \bibinfo {author} {\bibfnamefont {L.}~\bibnamefont {Li}}, \bibinfo {author} {\bibfnamefont {M.}~\bibnamefont {Stone}}, \bibinfo {author} {\bibfnamefont {G.}~\bibnamefont {Granroth}}, \bibinfo {author} {\bibfnamefont {M.}~\bibnamefont {Lumsden}}, \bibinfo {author} {\bibfnamefont {Y.}~\bibnamefont {Yiu}}, \bibinfo {author} {\bibfnamefont {J.}~\bibnamefont {Knolle}}, \emph {et~al.},\ }\bibfield  {title} {\bibinfo {title} {Proximate kitaev quantum spin liquid behaviour in a honeycomb magnet},\ }\href@noop {} {\bibfield  {journal} {\bibinfo  {journal} {Nature materials}\ }\textbf {\bibinfo {volume} {15}},\ \bibinfo {pages} {733} (\bibinfo {year} {2016})}\BibitemShut {NoStop}%
\bibitem [{\citenamefont {Kasahara}\ \emph {et~al.}(2018)\citenamefont {Kasahara}, \citenamefont {Ohnishi}, \citenamefont {Mizukami}, \citenamefont {Tanaka}, \citenamefont {Ma}, \citenamefont {Sugii}, \citenamefont {Kurita}, \citenamefont {Tanaka}, \citenamefont {Nasu}, \citenamefont {Motome} \emph {et~al.}}]{kasahara2018majorana}%
  \BibitemOpen
  \bibfield  {author} {\bibinfo {author} {\bibfnamefont {Y.}~\bibnamefont {Kasahara}}, \bibinfo {author} {\bibfnamefont {T.}~\bibnamefont {Ohnishi}}, \bibinfo {author} {\bibfnamefont {Y.}~\bibnamefont {Mizukami}}, \bibinfo {author} {\bibfnamefont {O.}~\bibnamefont {Tanaka}}, \bibinfo {author} {\bibfnamefont {S.}~\bibnamefont {Ma}}, \bibinfo {author} {\bibfnamefont {K.}~\bibnamefont {Sugii}}, \bibinfo {author} {\bibfnamefont {N.}~\bibnamefont {Kurita}}, \bibinfo {author} {\bibfnamefont {H.}~\bibnamefont {Tanaka}}, \bibinfo {author} {\bibfnamefont {J.}~\bibnamefont {Nasu}}, \bibinfo {author} {\bibfnamefont {Y.}~\bibnamefont {Motome}}, \emph {et~al.},\ }\bibfield  {title} {\bibinfo {title} {Majorana quantization and half-integer thermal quantum hall effect in a kitaev spin liquid},\ }\href@noop {} {\bibfield  {journal} {\bibinfo  {journal} {Nature}\ }\textbf {\bibinfo {volume} {559}},\ \bibinfo {pages} {227} (\bibinfo {year} {2018})}\BibitemShut {NoStop}%
\bibitem [{\citenamefont {Yokoi}\ \emph {et~al.}(2021)\citenamefont {Yokoi}, \citenamefont {Ma}, \citenamefont {Kasahara}, \citenamefont {Kasahara}, \citenamefont {Shibauchi}, \citenamefont {Kurita}, \citenamefont {Tanaka}, \citenamefont {Nasu}, \citenamefont {Motome}, \citenamefont {Hickey} \emph {et~al.}}]{yokoi2021quantum}%
  \BibitemOpen
  \bibfield  {author} {\bibinfo {author} {\bibfnamefont {T.}~\bibnamefont {Yokoi}}, \bibinfo {author} {\bibfnamefont {S.}~\bibnamefont {Ma}}, \bibinfo {author} {\bibfnamefont {Y.}~\bibnamefont {Kasahara}}, \bibinfo {author} {\bibfnamefont {S.}~\bibnamefont {Kasahara}}, \bibinfo {author} {\bibfnamefont {T.}~\bibnamefont {Shibauchi}}, \bibinfo {author} {\bibfnamefont {N.}~\bibnamefont {Kurita}}, \bibinfo {author} {\bibfnamefont {H.}~\bibnamefont {Tanaka}}, \bibinfo {author} {\bibfnamefont {J.}~\bibnamefont {Nasu}}, \bibinfo {author} {\bibfnamefont {Y.}~\bibnamefont {Motome}}, \bibinfo {author} {\bibfnamefont {C.}~\bibnamefont {Hickey}}, \emph {et~al.},\ }\bibfield  {title} {\bibinfo {title} {Quantum criticality and emergent dirac fermions in a frustrated magnet},\ }\href@noop {} {\bibfield  {journal} {\bibinfo  {journal} {Science}\ }\textbf {\bibinfo {volume} {373}},\ \bibinfo {pages} {568} (\bibinfo {year} {2021})}\BibitemShut {NoStop}%
\bibitem [{\citenamefont {Trebst}\ and\ \citenamefont {Hickey}(2022)}]{trebst2022kitaev}%
  \BibitemOpen
  \bibfield  {author} {\bibinfo {author} {\bibfnamefont {S.}~\bibnamefont {Trebst}}\ and\ \bibinfo {author} {\bibfnamefont {C.}~\bibnamefont {Hickey}},\ }\bibfield  {title} {\bibinfo {title} {The kitaev quantum spin liquid: a review},\ }\href@noop {} {\bibfield  {journal} {\bibinfo  {journal} {Physics Reports}\ }\textbf {\bibinfo {volume} {950}},\ \bibinfo {pages} {1} (\bibinfo {year} {2022})}\BibitemShut {NoStop}%
\bibitem [{\citenamefont {Tokura}\ \emph {et~al.}(2017)\citenamefont {Tokura}, \citenamefont {Kawasaki},\ and\ \citenamefont {Nagaosa}}]{tokura2017emergent}%
  \BibitemOpen
  \bibfield  {author} {\bibinfo {author} {\bibfnamefont {Y.}~\bibnamefont {Tokura}}, \bibinfo {author} {\bibfnamefont {M.}~\bibnamefont {Kawasaki}},\ and\ \bibinfo {author} {\bibfnamefont {N.}~\bibnamefont {Nagaosa}},\ }\bibfield  {title} {\bibinfo {title} {Emergent functions of quantum materials},\ }\href@noop {} {\bibfield  {journal} {\bibinfo  {journal} {Nature Physics}\ }\textbf {\bibinfo {volume} {13}},\ \bibinfo {pages} {1056} (\bibinfo {year} {2017})}\BibitemShut {NoStop}%
\bibitem [{\citenamefont {Norman}(2016)}]{norman2016colloquium}%
  \BibitemOpen
  \bibfield  {author} {\bibinfo {author} {\bibfnamefont {M.}~\bibnamefont {Norman}},\ }\bibfield  {title} {\bibinfo {title} {Colloquium: Herbertsmithite and the search for the quantum spin liquid},\ }\href@noop {} {\bibfield  {journal} {\bibinfo  {journal} {Reviews of Modern Physics}\ }\textbf {\bibinfo {volume} {88}},\ \bibinfo {pages} {041002} (\bibinfo {year} {2016})}\BibitemShut {NoStop}%
\bibitem [{\citenamefont {Wen}\ \emph {et~al.}(2019)\citenamefont {Wen}, \citenamefont {Yu}, \citenamefont {Li}, \citenamefont {Yu},\ and\ \citenamefont {Li}}]{wen2019}%
  \BibitemOpen
  \bibfield  {author} {\bibinfo {author} {\bibfnamefont {J.}~\bibnamefont {Wen}}, \bibinfo {author} {\bibfnamefont {S.-L.}\ \bibnamefont {Yu}}, \bibinfo {author} {\bibfnamefont {S.}~\bibnamefont {Li}}, \bibinfo {author} {\bibfnamefont {W.}~\bibnamefont {Yu}},\ and\ \bibinfo {author} {\bibfnamefont {J.-X.}\ \bibnamefont {Li}},\ }\bibfield  {title} {\bibinfo {title} {Experimental identification of quantum spin liquids},\ }\href@noop {} {\bibfield  {journal} {\bibinfo  {journal} {npj Quantum Materials}\ }\textbf {\bibinfo {volume} {4}},\ \bibinfo {pages} {12} (\bibinfo {year} {2019})}\BibitemShut {NoStop}%
\bibitem [{\citenamefont {Clark}\ and\ \citenamefont {Abdeldaim}(2021)}]{clark2021}%
  \BibitemOpen
  \bibfield  {author} {\bibinfo {author} {\bibfnamefont {L.}~\bibnamefont {Clark}}\ and\ \bibinfo {author} {\bibfnamefont {A.~H.}\ \bibnamefont {Abdeldaim}},\ }\bibfield  {title} {\bibinfo {title} {Quantum spin liquids from a materials perspective},\ }\href@noop {} {\bibfield  {journal} {\bibinfo  {journal} {Annual Review of Materials Research}\ }\textbf {\bibinfo {volume} {51}},\ \bibinfo {pages} {495} (\bibinfo {year} {2021})}\BibitemShut {NoStop}%
\bibitem [{\citenamefont {Ross}\ \emph {et~al.}(2009)\citenamefont {Ross}, \citenamefont {Ruff}, \citenamefont {Adams}, \citenamefont {Gardner}, \citenamefont {Dabkowska}, \citenamefont {Qiu}, \citenamefont {Copley},\ and\ \citenamefont {Gaulin}}]{ross2009}%
  \BibitemOpen
  \bibfield  {author} {\bibinfo {author} {\bibfnamefont {K.}~\bibnamefont {Ross}}, \bibinfo {author} {\bibfnamefont {J.}~\bibnamefont {Ruff}}, \bibinfo {author} {\bibfnamefont {C.}~\bibnamefont {Adams}}, \bibinfo {author} {\bibfnamefont {J.}~\bibnamefont {Gardner}}, \bibinfo {author} {\bibfnamefont {H.}~\bibnamefont {Dabkowska}}, \bibinfo {author} {\bibfnamefont {Y.}~\bibnamefont {Qiu}}, \bibinfo {author} {\bibfnamefont {J.}~\bibnamefont {Copley}},\ and\ \bibinfo {author} {\bibfnamefont {B.}~\bibnamefont {Gaulin}},\ }\bibfield  {title} {\bibinfo {title} {Two-dimensional kagome correlations and field induced order in the ferromagnetic \ch{XY} pyrochlore \ch{Yb2Ti2O7}},\ }\href@noop {} {\bibfield  {journal} {\bibinfo  {journal} {Physical Review Letters}\ }\textbf {\bibinfo {volume} {103}},\ \bibinfo {pages} {227202} (\bibinfo {year} {2009})}\BibitemShut {NoStop}%
\bibitem [{\citenamefont {Thompson}\ \emph {et~al.}(2011)\citenamefont {Thompson}, \citenamefont {McClarty}, \citenamefont {R{\o}nnow}, \citenamefont {Regnault}, \citenamefont {Sorge},\ and\ \citenamefont {Gingras}}]{thompson2011}%
  \BibitemOpen
  \bibfield  {author} {\bibinfo {author} {\bibfnamefont {J.~D.}\ \bibnamefont {Thompson}}, \bibinfo {author} {\bibfnamefont {P.~A.}\ \bibnamefont {McClarty}}, \bibinfo {author} {\bibfnamefont {H.~M.}\ \bibnamefont {R{\o}nnow}}, \bibinfo {author} {\bibfnamefont {L.~P.}\ \bibnamefont {Regnault}}, \bibinfo {author} {\bibfnamefont {A.}~\bibnamefont {Sorge}},\ and\ \bibinfo {author} {\bibfnamefont {M.~J.}\ \bibnamefont {Gingras}},\ }\bibfield  {title} {\bibinfo {title} {Rods of neutron scattering intensity in \ch{Yb2Ti2O7}: compelling evidence for significant anisotropic exchange in a magnetic pyrochlore oxide},\ }\href@noop {} {\bibfield  {journal} {\bibinfo  {journal} {Physical Review Letters}\ }\textbf {\bibinfo {volume} {106}},\ \bibinfo {pages} {187202} (\bibinfo {year} {2011})}\BibitemShut {NoStop}%
\bibitem [{\citenamefont {Yaouanc}\ \emph {et~al.}(2011)\citenamefont {Yaouanc}, \citenamefont {De~R{\'e}otier}, \citenamefont {Marin},\ and\ \citenamefont {Glazkov}}]{yaouanc2011}%
  \BibitemOpen
  \bibfield  {author} {\bibinfo {author} {\bibfnamefont {A.}~\bibnamefont {Yaouanc}}, \bibinfo {author} {\bibfnamefont {P.~D.}\ \bibnamefont {De~R{\'e}otier}}, \bibinfo {author} {\bibfnamefont {C.}~\bibnamefont {Marin}},\ and\ \bibinfo {author} {\bibfnamefont {V.}~\bibnamefont {Glazkov}},\ }\bibfield  {title} {\bibinfo {title} {Single-crystal versus polycrystalline samples of magnetically frustrated \ch{Yb2Ti2O7}: specific heat results},\ }\href@noop {} {\bibfield  {journal} {\bibinfo  {journal} {Physical Review B}\ }\textbf {\bibinfo {volume} {84}},\ \bibinfo {pages} {172408} (\bibinfo {year} {2011})}\BibitemShut {NoStop}%
\bibitem [{\citenamefont {Chang}\ \emph {et~al.}(2012)\citenamefont {Chang}, \citenamefont {Onoda}, \citenamefont {Su}, \citenamefont {Kao}, \citenamefont {Tsuei}, \citenamefont {Yasui}, \citenamefont {Kakurai},\ and\ \citenamefont {Lees}}]{Chang2012}%
  \BibitemOpen
  \bibfield  {author} {\bibinfo {author} {\bibfnamefont {L.-J.}\ \bibnamefont {Chang}}, \bibinfo {author} {\bibfnamefont {S.}~\bibnamefont {Onoda}}, \bibinfo {author} {\bibfnamefont {Y.}~\bibnamefont {Su}}, \bibinfo {author} {\bibfnamefont {Y.-J.}\ \bibnamefont {Kao}}, \bibinfo {author} {\bibfnamefont {K.-D.}\ \bibnamefont {Tsuei}}, \bibinfo {author} {\bibfnamefont {Y.}~\bibnamefont {Yasui}}, \bibinfo {author} {\bibfnamefont {K.}~\bibnamefont {Kakurai}},\ and\ \bibinfo {author} {\bibfnamefont {M.~R.}\ \bibnamefont {Lees}},\ }\bibfield  {title} {\bibinfo {title} {Higgs transition from a magnetic coulomb liquid to a ferromagnet in \ch{Yb2Ti2O7}},\ }\href@noop {} {\bibfield  {journal} {\bibinfo  {journal} {Nature communications}\ }\textbf {\bibinfo {volume} {3}},\ \bibinfo {pages} {992} (\bibinfo {year} {2012})}\BibitemShut {NoStop}%
\bibitem [{\citenamefont {D’Ortenzio}\ \emph {et~al.}(2013)\citenamefont {D’Ortenzio}, \citenamefont {Dabkowska}, \citenamefont {Dunsiger}, \citenamefont {Gaulin}, \citenamefont {Gingras}, \citenamefont {Goko}, \citenamefont {Kycia}, \citenamefont {Liu}, \citenamefont {Medina}, \citenamefont {Munsie} \emph {et~al.}}]{DOrtenzio2013}%
  \BibitemOpen
  \bibfield  {author} {\bibinfo {author} {\bibfnamefont {R.}~\bibnamefont {D’Ortenzio}}, \bibinfo {author} {\bibfnamefont {H.}~\bibnamefont {Dabkowska}}, \bibinfo {author} {\bibfnamefont {S.}~\bibnamefont {Dunsiger}}, \bibinfo {author} {\bibfnamefont {B.}~\bibnamefont {Gaulin}}, \bibinfo {author} {\bibfnamefont {M.}~\bibnamefont {Gingras}}, \bibinfo {author} {\bibfnamefont {T.}~\bibnamefont {Goko}}, \bibinfo {author} {\bibfnamefont {J.}~\bibnamefont {Kycia}}, \bibinfo {author} {\bibfnamefont {L.}~\bibnamefont {Liu}}, \bibinfo {author} {\bibfnamefont {T.}~\bibnamefont {Medina}}, \bibinfo {author} {\bibfnamefont {T.}~\bibnamefont {Munsie}}, \emph {et~al.},\ }\bibfield  {title} {\bibinfo {title} {Unconventional magnetic ground state in \ch{Yb2Ti2O7}},\ }\href@noop {} {\bibfield  {journal} {\bibinfo  {journal} {Physical Review B}\ }\textbf {\bibinfo {volume} {88}},\ \bibinfo {pages} {134428} (\bibinfo {year} {2013})}\BibitemShut {NoStop}%
\bibitem [{\citenamefont {Lhotel}\ \emph {et~al.}(2014)\citenamefont {Lhotel}, \citenamefont {Giblin}, \citenamefont {Lees}, \citenamefont {Balakrishnan}, \citenamefont {Chang},\ and\ \citenamefont {Yasui}}]{Lhotel2014}%
  \BibitemOpen
  \bibfield  {author} {\bibinfo {author} {\bibfnamefont {E.}~\bibnamefont {Lhotel}}, \bibinfo {author} {\bibfnamefont {S.}~\bibnamefont {Giblin}}, \bibinfo {author} {\bibfnamefont {M.~R.}\ \bibnamefont {Lees}}, \bibinfo {author} {\bibfnamefont {G.}~\bibnamefont {Balakrishnan}}, \bibinfo {author} {\bibfnamefont {L.}~\bibnamefont {Chang}},\ and\ \bibinfo {author} {\bibfnamefont {Y.}~\bibnamefont {Yasui}},\ }\bibfield  {title} {\bibinfo {title} {First-order magnetic transition in \ch{Yb2Ti2O7}},\ }\href@noop {} {\bibfield  {journal} {\bibinfo  {journal} {Physical Review B}\ }\textbf {\bibinfo {volume} {89}},\ \bibinfo {pages} {224419} (\bibinfo {year} {2014})}\BibitemShut {NoStop}%
\bibitem [{\citenamefont {Thompson}\ \emph {et~al.}(2017)\citenamefont {Thompson}, \citenamefont {McClarty}, \citenamefont {Prabhakaran}, \citenamefont {Cabrera}, \citenamefont {Guidi},\ and\ \citenamefont {Coldea}}]{Thompson2017}%
  \BibitemOpen
  \bibfield  {author} {\bibinfo {author} {\bibfnamefont {J.}~\bibnamefont {Thompson}}, \bibinfo {author} {\bibfnamefont {P.~A.}\ \bibnamefont {McClarty}}, \bibinfo {author} {\bibfnamefont {D.}~\bibnamefont {Prabhakaran}}, \bibinfo {author} {\bibfnamefont {I.}~\bibnamefont {Cabrera}}, \bibinfo {author} {\bibfnamefont {T.}~\bibnamefont {Guidi}},\ and\ \bibinfo {author} {\bibfnamefont {R.}~\bibnamefont {Coldea}},\ }\bibfield  {title} {\bibinfo {title} {Quasiparticle breakdown and spin hamiltonian of the frustrated quantum pyrochlore \ch{Yb2Ti2O7} in a magnetic field},\ }\href@noop {} {\bibfield  {journal} {\bibinfo  {journal} {Physical Review Letters}\ }\textbf {\bibinfo {volume} {119}},\ \bibinfo {pages} {057203} (\bibinfo {year} {2017})}\BibitemShut {NoStop}%
\bibitem [{\citenamefont {Ross}\ \emph {et~al.}(2011)\citenamefont {Ross}, \citenamefont {Savary}, \citenamefont {Gaulin},\ and\ \citenamefont {Balents}}]{ross2011}%
  \BibitemOpen
  \bibfield  {author} {\bibinfo {author} {\bibfnamefont {K.~A.}\ \bibnamefont {Ross}}, \bibinfo {author} {\bibfnamefont {L.}~\bibnamefont {Savary}}, \bibinfo {author} {\bibfnamefont {B.~D.}\ \bibnamefont {Gaulin}},\ and\ \bibinfo {author} {\bibfnamefont {L.}~\bibnamefont {Balents}},\ }\bibfield  {title} {\bibinfo {title} {Quantum excitations in quantum spin ice},\ }\href@noop {} {\bibfield  {journal} {\bibinfo  {journal} {Physical Review X}\ }\textbf {\bibinfo {volume} {1}},\ \bibinfo {pages} {021002} (\bibinfo {year} {2011})}\BibitemShut {NoStop}%
\bibitem [{\citenamefont {Robert}\ \emph {et~al.}(2015)\citenamefont {Robert}, \citenamefont {Lhotel}, \citenamefont {Remenyi}, \citenamefont {Sahling}, \citenamefont {Mirebeau}, \citenamefont {Decorse}, \citenamefont {Canals},\ and\ \citenamefont {Petit}}]{Robert2015}%
  \BibitemOpen
  \bibfield  {author} {\bibinfo {author} {\bibfnamefont {J.}~\bibnamefont {Robert}}, \bibinfo {author} {\bibfnamefont {E.}~\bibnamefont {Lhotel}}, \bibinfo {author} {\bibfnamefont {G.}~\bibnamefont {Remenyi}}, \bibinfo {author} {\bibfnamefont {S.}~\bibnamefont {Sahling}}, \bibinfo {author} {\bibfnamefont {I.}~\bibnamefont {Mirebeau}}, \bibinfo {author} {\bibfnamefont {C.}~\bibnamefont {Decorse}}, \bibinfo {author} {\bibfnamefont {B.}~\bibnamefont {Canals}},\ and\ \bibinfo {author} {\bibfnamefont {S.}~\bibnamefont {Petit}},\ }\bibfield  {title} {\bibinfo {title} {Spin dynamics in the presence of competing ferromagnetic and antiferromagnetic correlations in \ch{Yb2Ti2O7}},\ }\href@noop {} {\bibfield  {journal} {\bibinfo  {journal} {Physical Review B}\ }\textbf {\bibinfo {volume} {92}},\ \bibinfo {pages} {064425} (\bibinfo {year} {2015})}\BibitemShut {NoStop}%
\bibitem [{\citenamefont {Gaudet}\ \emph {et~al.}(2016)\citenamefont {Gaudet}, \citenamefont {Ross}, \citenamefont {Kermarrec}, \citenamefont {Butch}, \citenamefont {Ehlers}, \citenamefont {Dabkowska},\ and\ \citenamefont {Gaulin}}]{Gaudet2016}%
  \BibitemOpen
  \bibfield  {author} {\bibinfo {author} {\bibfnamefont {J.}~\bibnamefont {Gaudet}}, \bibinfo {author} {\bibfnamefont {K.}~\bibnamefont {Ross}}, \bibinfo {author} {\bibfnamefont {E.}~\bibnamefont {Kermarrec}}, \bibinfo {author} {\bibfnamefont {N.}~\bibnamefont {Butch}}, \bibinfo {author} {\bibfnamefont {G.}~\bibnamefont {Ehlers}}, \bibinfo {author} {\bibfnamefont {H.}~\bibnamefont {Dabkowska}},\ and\ \bibinfo {author} {\bibfnamefont {B.}~\bibnamefont {Gaulin}},\ }\bibfield  {title} {\bibinfo {title} {Gapless quantum excitations from an icelike splayed ferromagnetic ground state in stoichiometric \ch{Yb2Ti2O7}},\ }\href@noop {} {\bibfield  {journal} {\bibinfo  {journal} {Physical Review B}\ }\textbf {\bibinfo {volume} {93}},\ \bibinfo {pages} {064406} (\bibinfo {year} {2016})}\BibitemShut {NoStop}%
\bibitem [{\citenamefont {Pan}\ \emph {et~al.}(2016)\citenamefont {Pan}, \citenamefont {Laurita}, \citenamefont {Ross}, \citenamefont {Gaulin},\ and\ \citenamefont {Armitage}}]{Pan2016}%
  \BibitemOpen
  \bibfield  {author} {\bibinfo {author} {\bibfnamefont {L.}~\bibnamefont {Pan}}, \bibinfo {author} {\bibfnamefont {N.}~\bibnamefont {Laurita}}, \bibinfo {author} {\bibfnamefont {K.~A.}\ \bibnamefont {Ross}}, \bibinfo {author} {\bibfnamefont {B.~D.}\ \bibnamefont {Gaulin}},\ and\ \bibinfo {author} {\bibfnamefont {N.}~\bibnamefont {Armitage}},\ }\bibfield  {title} {\bibinfo {title} {A measure of monopole inertia in the quantum spin ice \ch{Yb2Ti2O7}},\ }\href@noop {} {\bibfield  {journal} {\bibinfo  {journal} {Nature Physics}\ }\textbf {\bibinfo {volume} {12}},\ \bibinfo {pages} {361} (\bibinfo {year} {2016})}\BibitemShut {NoStop}%
\bibitem [{\citenamefont {Pe{\c{c}}anha-Antonio}\ \emph {et~al.}(2017)\citenamefont {Pe{\c{c}}anha-Antonio}, \citenamefont {Feng}, \citenamefont {Su}, \citenamefont {Pomjakushin}, \citenamefont {Demmel}, \citenamefont {Chang}, \citenamefont {Aldus}, \citenamefont {Xiao}, \citenamefont {Lees},\ and\ \citenamefont {Br{\"u}ckel}}]{PecanhaAntonio2017}%
  \BibitemOpen
  \bibfield  {author} {\bibinfo {author} {\bibfnamefont {V.}~\bibnamefont {Pe{\c{c}}anha-Antonio}}, \bibinfo {author} {\bibfnamefont {E.}~\bibnamefont {Feng}}, \bibinfo {author} {\bibfnamefont {Y.}~\bibnamefont {Su}}, \bibinfo {author} {\bibfnamefont {V.}~\bibnamefont {Pomjakushin}}, \bibinfo {author} {\bibfnamefont {F.}~\bibnamefont {Demmel}}, \bibinfo {author} {\bibfnamefont {L.-J.}\ \bibnamefont {Chang}}, \bibinfo {author} {\bibfnamefont {R.~J.}\ \bibnamefont {Aldus}}, \bibinfo {author} {\bibfnamefont {Y.}~\bibnamefont {Xiao}}, \bibinfo {author} {\bibfnamefont {M.~R.}\ \bibnamefont {Lees}},\ and\ \bibinfo {author} {\bibfnamefont {T.}~\bibnamefont {Br{\"u}ckel}},\ }\bibfield  {title} {\bibinfo {title} {Magnetic excitations in the ground state of \ch{Yb2Ti2O7}},\ }\href@noop {} {\bibfield  {journal} {\bibinfo  {journal} {Physical Review B}\ }\textbf {\bibinfo {volume} {96}},\ \bibinfo {pages} {214415} (\bibinfo {year} {2017})}\BibitemShut {NoStop}%
\bibitem [{\citenamefont {Arpino}\ \emph {et~al.}(2017)\citenamefont {Arpino}, \citenamefont {Trump}, \citenamefont {Scheie}, \citenamefont {McQueen},\ and\ \citenamefont {Koohpayeh}}]{Arpino2017}%
  \BibitemOpen
  \bibfield  {author} {\bibinfo {author} {\bibfnamefont {K.}~\bibnamefont {Arpino}}, \bibinfo {author} {\bibfnamefont {B.}~\bibnamefont {Trump}}, \bibinfo {author} {\bibfnamefont {A.}~\bibnamefont {Scheie}}, \bibinfo {author} {\bibfnamefont {T.}~\bibnamefont {McQueen}},\ and\ \bibinfo {author} {\bibfnamefont {S.}~\bibnamefont {Koohpayeh}},\ }\bibfield  {title} {\bibinfo {title} {Impact of stoichiometry of \ch{Yb2Ti2O7} on its physical properties},\ }\href@noop {} {\bibfield  {journal} {\bibinfo  {journal} {Physical Review B}\ }\textbf {\bibinfo {volume} {95}},\ \bibinfo {pages} {094407} (\bibinfo {year} {2017})}\BibitemShut {NoStop}%
\bibitem [{\citenamefont {Scheie}\ \emph {et~al.}(2017)\citenamefont {Scheie}, \citenamefont {Kindervater}, \citenamefont {S{\"a}ubert}, \citenamefont {Duvinage}, \citenamefont {Pfleiderer}, \citenamefont {Changlani}, \citenamefont {Zhang}, \citenamefont {Harriger}, \citenamefont {Arpino}, \citenamefont {Koohpayeh} \emph {et~al.}}]{scheie2017reentrant}%
  \BibitemOpen
  \bibfield  {author} {\bibinfo {author} {\bibfnamefont {A.}~\bibnamefont {Scheie}}, \bibinfo {author} {\bibfnamefont {J.}~\bibnamefont {Kindervater}}, \bibinfo {author} {\bibfnamefont {S.}~\bibnamefont {S{\"a}ubert}}, \bibinfo {author} {\bibfnamefont {C.}~\bibnamefont {Duvinage}}, \bibinfo {author} {\bibfnamefont {C.}~\bibnamefont {Pfleiderer}}, \bibinfo {author} {\bibfnamefont {H.}~\bibnamefont {Changlani}}, \bibinfo {author} {\bibfnamefont {S.}~\bibnamefont {Zhang}}, \bibinfo {author} {\bibfnamefont {L.}~\bibnamefont {Harriger}}, \bibinfo {author} {\bibfnamefont {K.}~\bibnamefont {Arpino}}, \bibinfo {author} {\bibfnamefont {S.}~\bibnamefont {Koohpayeh}}, \emph {et~al.},\ }\bibfield  {title} {\bibinfo {title} {Reentrant phase diagram of \ch{Yb2Ti2O7} in a \textless111\textgreater magnetic field},\ }\href@noop {} {\bibfield  {journal} {\bibinfo  {journal} {Physical Review Letters}\ }\textbf {\bibinfo {volume} {119}},\ \bibinfo {pages} {127201} (\bibinfo {year} {2017})}\BibitemShut {NoStop}%
\bibitem [{\citenamefont {Scheie}\ \emph {et~al.}(2020)\citenamefont {Scheie}, \citenamefont {Kindervater}, \citenamefont {Zhang}, \citenamefont {Changlani}, \citenamefont {Sala}, \citenamefont {Ehlers}, \citenamefont {Heinemann}, \citenamefont {Tucker}, \citenamefont {Koohpayeh},\ and\ \citenamefont {Broholm}}]{Scheie2020a}%
  \BibitemOpen
  \bibfield  {author} {\bibinfo {author} {\bibfnamefont {A.}~\bibnamefont {Scheie}}, \bibinfo {author} {\bibfnamefont {J.}~\bibnamefont {Kindervater}}, \bibinfo {author} {\bibfnamefont {S.}~\bibnamefont {Zhang}}, \bibinfo {author} {\bibfnamefont {H.~J.}\ \bibnamefont {Changlani}}, \bibinfo {author} {\bibfnamefont {G.}~\bibnamefont {Sala}}, \bibinfo {author} {\bibfnamefont {G.}~\bibnamefont {Ehlers}}, \bibinfo {author} {\bibfnamefont {A.}~\bibnamefont {Heinemann}}, \bibinfo {author} {\bibfnamefont {G.~S.}\ \bibnamefont {Tucker}}, \bibinfo {author} {\bibfnamefont {S.~M.}\ \bibnamefont {Koohpayeh}},\ and\ \bibinfo {author} {\bibfnamefont {C.}~\bibnamefont {Broholm}},\ }\bibfield  {title} {\bibinfo {title} {Multiphase magnetism in \ch{Yb2Ti2O7}},\ }\href@noop {} {\bibfield  {journal} {\bibinfo  {journal} {Proceedings of the National Academy of Sciences}\ }\textbf {\bibinfo {volume} {117}},\ \bibinfo {pages} {27245} (\bibinfo {year} {2020})}\BibitemShut {NoStop}%
\bibitem [{\citenamefont {Petit}(2020)}]{petit2020way}%
  \BibitemOpen
  \bibfield  {author} {\bibinfo {author} {\bibfnamefont {S.}~\bibnamefont {Petit}},\ }\bibfield  {title} {\bibinfo {title} {On the way to understanding \ch{Yb2Ti2O7}},\ }\href@noop {} {\bibfield  {journal} {\bibinfo  {journal} {Proceedings of the National Academy of Sciences}\ }\textbf {\bibinfo {volume} {117}},\ \bibinfo {pages} {29263} (\bibinfo {year} {2020})}\BibitemShut {NoStop}%
\bibitem [{\citenamefont {Scheie}\ \emph {et~al.}(2022)\citenamefont {Scheie}, \citenamefont {Benton}, \citenamefont {Taillefumier}, \citenamefont {Jaubert}, \citenamefont {Sala}, \citenamefont {Jalarvo}, \citenamefont {Koohpayeh},\ and\ \citenamefont {Shannon}}]{scheie2022dynamical}%
  \BibitemOpen
  \bibfield  {author} {\bibinfo {author} {\bibfnamefont {A.}~\bibnamefont {Scheie}}, \bibinfo {author} {\bibfnamefont {O.}~\bibnamefont {Benton}}, \bibinfo {author} {\bibfnamefont {M.}~\bibnamefont {Taillefumier}}, \bibinfo {author} {\bibfnamefont {L.~D.}\ \bibnamefont {Jaubert}}, \bibinfo {author} {\bibfnamefont {G.}~\bibnamefont {Sala}}, \bibinfo {author} {\bibfnamefont {N.}~\bibnamefont {Jalarvo}}, \bibinfo {author} {\bibfnamefont {S.~M.}\ \bibnamefont {Koohpayeh}},\ and\ \bibinfo {author} {\bibfnamefont {N.}~\bibnamefont {Shannon}},\ }\bibfield  {title} {\bibinfo {title} {Dynamical scaling as a signature of multiple phase competition in \ch{Yb2Ti2O7}},\ }\href@noop {} {\bibfield  {journal} {\bibinfo  {journal} {Physical Review Letters}\ }\textbf {\bibinfo {volume} {129}},\ \bibinfo {pages} {217202} (\bibinfo {year} {2022})}\BibitemShut {NoStop}%
\bibitem [{\citenamefont {S{\"a}ubert}\ \emph {et~al.}(2020)\citenamefont {S{\"a}ubert}, \citenamefont {Scheie}, \citenamefont {Duvinage}, \citenamefont {Kindervater}, \citenamefont {Zhang}, \citenamefont {Changlani}, \citenamefont {Xu}, \citenamefont {Koohpayeh}, \citenamefont {Tchernyshyov}, \citenamefont {Broholm} \emph {et~al.}}]{saubert2020}%
  \BibitemOpen
  \bibfield  {author} {\bibinfo {author} {\bibfnamefont {S.}~\bibnamefont {S{\"a}ubert}}, \bibinfo {author} {\bibfnamefont {A.}~\bibnamefont {Scheie}}, \bibinfo {author} {\bibfnamefont {C.}~\bibnamefont {Duvinage}}, \bibinfo {author} {\bibfnamefont {J.}~\bibnamefont {Kindervater}}, \bibinfo {author} {\bibfnamefont {S.}~\bibnamefont {Zhang}}, \bibinfo {author} {\bibfnamefont {H.}~\bibnamefont {Changlani}}, \bibinfo {author} {\bibfnamefont {G.}~\bibnamefont {Xu}}, \bibinfo {author} {\bibfnamefont {S.}~\bibnamefont {Koohpayeh}}, \bibinfo {author} {\bibfnamefont {O.}~\bibnamefont {Tchernyshyov}}, \bibinfo {author} {\bibfnamefont {C.~L.}\ \bibnamefont {Broholm}}, \emph {et~al.},\ }\bibfield  {title} {\bibinfo {title} {Orientation dependence of the magnetic phase diagram of \ch{Yb2Ti2O7}},\ }\href@noop {} {\bibfield  {journal} {\bibinfo  {journal} {Physical Review B}\ }\textbf {\bibinfo {volume} {101}},\ \bibinfo {pages} {174434} (\bibinfo {year} {2020})}\BibitemShut {NoStop}%
\bibitem [{\citenamefont {Zhang}\ \emph {et~al.}(2023)\citenamefont {Zhang}, \citenamefont {Xing}, \citenamefont {Noordhoek}, \citenamefont {Liu}, \citenamefont {Zhao}, \citenamefont {Hor{\'a}k}, \citenamefont {Huang}, \citenamefont {Hao}, \citenamefont {Yang}, \citenamefont {Pandey} \emph {et~al.}}]{zhang2023anomalous}%
  \BibitemOpen
  \bibfield  {author} {\bibinfo {author} {\bibfnamefont {H.}~\bibnamefont {Zhang}}, \bibinfo {author} {\bibfnamefont {C.}~\bibnamefont {Xing}}, \bibinfo {author} {\bibfnamefont {K.}~\bibnamefont {Noordhoek}}, \bibinfo {author} {\bibfnamefont {Z.}~\bibnamefont {Liu}}, \bibinfo {author} {\bibfnamefont {T.}~\bibnamefont {Zhao}}, \bibinfo {author} {\bibfnamefont {L.}~\bibnamefont {Hor{\'a}k}}, \bibinfo {author} {\bibfnamefont {Q.}~\bibnamefont {Huang}}, \bibinfo {author} {\bibfnamefont {L.}~\bibnamefont {Hao}}, \bibinfo {author} {\bibfnamefont {J.}~\bibnamefont {Yang}}, \bibinfo {author} {\bibfnamefont {S.}~\bibnamefont {Pandey}}, \emph {et~al.},\ }\bibfield  {title} {\bibinfo {title} {Anomalous magnetoresistance by breaking ice rule in \ch{Bi2Ir2O7}/\ch{Dy2Ti2O7} heterostructure},\ }\href@noop {} {\bibfield  {journal} {\bibinfo  {journal} {Nature Communications}\ }\textbf {\bibinfo {volume} {14}},\ \bibinfo {pages} {1404} (\bibinfo {year} {2023})}\BibitemShut {NoStop}%
\bibitem [{\citenamefont {Bramwell}\ and\ \citenamefont {Gingras}(2001)}]{bramwell2001spin}%
  \BibitemOpen
  \bibfield  {author} {\bibinfo {author} {\bibfnamefont {S.~T.}\ \bibnamefont {Bramwell}}\ and\ \bibinfo {author} {\bibfnamefont {M.~J.}\ \bibnamefont {Gingras}},\ }\bibfield  {title} {\bibinfo {title} {Spin ice state in frustrated magnetic pyrochlore materials},\ }\href@noop {} {\bibfield  {journal} {\bibinfo  {journal} {Science}\ }\textbf {\bibinfo {volume} {294}},\ \bibinfo {pages} {1495} (\bibinfo {year} {2001})}\BibitemShut {NoStop}%
\bibitem [{\citenamefont {Siddharthan}\ \emph {et~al.}(1999)\citenamefont {Siddharthan}, \citenamefont {Shastry}, \citenamefont {Ramirez}, \citenamefont {Hayashi}, \citenamefont {Cava},\ and\ \citenamefont {Rosenkranz}}]{siddharthan1999ising}%
  \BibitemOpen
  \bibfield  {author} {\bibinfo {author} {\bibfnamefont {R.}~\bibnamefont {Siddharthan}}, \bibinfo {author} {\bibfnamefont {B.}~\bibnamefont {Shastry}}, \bibinfo {author} {\bibfnamefont {A.}~\bibnamefont {Ramirez}}, \bibinfo {author} {\bibfnamefont {A.}~\bibnamefont {Hayashi}}, \bibinfo {author} {\bibfnamefont {R.}~\bibnamefont {Cava}},\ and\ \bibinfo {author} {\bibfnamefont {S.}~\bibnamefont {Rosenkranz}},\ }\bibfield  {title} {\bibinfo {title} {Ising pyrochlore magnets: Low-temperature properties,“ice rules,” and beyond},\ }\href@noop {} {\bibfield  {journal} {\bibinfo  {journal} {Physical Review Letters}\ }\textbf {\bibinfo {volume} {83}},\ \bibinfo {pages} {1854} (\bibinfo {year} {1999})}\BibitemShut {NoStop}%
\bibitem [{\citenamefont {Ramirez}\ \emph {et~al.}(1999)\citenamefont {Ramirez}, \citenamefont {Hayashi}, \citenamefont {Cava}, \citenamefont {Siddharthan},\ and\ \citenamefont {Shastry}}]{ramirez1999zero}%
  \BibitemOpen
  \bibfield  {author} {\bibinfo {author} {\bibfnamefont {A.~P.}\ \bibnamefont {Ramirez}}, \bibinfo {author} {\bibfnamefont {A.}~\bibnamefont {Hayashi}}, \bibinfo {author} {\bibfnamefont {R.~J.}\ \bibnamefont {Cava}}, \bibinfo {author} {\bibfnamefont {R.}~\bibnamefont {Siddharthan}},\ and\ \bibinfo {author} {\bibfnamefont {B.}~\bibnamefont {Shastry}},\ }\bibfield  {title} {\bibinfo {title} {Zero-point entropy in ‘spin ice’},\ }\href@noop {} {\bibfield  {journal} {\bibinfo  {journal} {Nature}\ }\textbf {\bibinfo {volume} {399}},\ \bibinfo {pages} {333} (\bibinfo {year} {1999})}\BibitemShut {NoStop}%
\bibitem [{\citenamefont {Castelnovo}\ \emph {et~al.}(2008)\citenamefont {Castelnovo}, \citenamefont {Moessner},\ and\ \citenamefont {Sondhi}}]{castelnovo2008magnetic}%
  \BibitemOpen
  \bibfield  {author} {\bibinfo {author} {\bibfnamefont {C.}~\bibnamefont {Castelnovo}}, \bibinfo {author} {\bibfnamefont {R.}~\bibnamefont {Moessner}},\ and\ \bibinfo {author} {\bibfnamefont {S.~L.}\ \bibnamefont {Sondhi}},\ }\bibfield  {title} {\bibinfo {title} {Magnetic monopoles in spin ice},\ }\href@noop {} {\bibfield  {journal} {\bibinfo  {journal} {Nature}\ }\textbf {\bibinfo {volume} {451}},\ \bibinfo {pages} {42} (\bibinfo {year} {2008})}\BibitemShut {NoStop}%
\bibitem [{\citenamefont {Tabata}\ \emph {et~al.}(2006)\citenamefont {Tabata}, \citenamefont {Kadowaki}, \citenamefont {Matsuhira}, \citenamefont {Hiroi}, \citenamefont {Aso}, \citenamefont {Ressouche},\ and\ \citenamefont {F{\aa}k}}]{tabata2006kagome}%
  \BibitemOpen
  \bibfield  {author} {\bibinfo {author} {\bibfnamefont {Y.}~\bibnamefont {Tabata}}, \bibinfo {author} {\bibfnamefont {H.}~\bibnamefont {Kadowaki}}, \bibinfo {author} {\bibfnamefont {K.}~\bibnamefont {Matsuhira}}, \bibinfo {author} {\bibfnamefont {Z.}~\bibnamefont {Hiroi}}, \bibinfo {author} {\bibfnamefont {N.}~\bibnamefont {Aso}}, \bibinfo {author} {\bibfnamefont {E.}~\bibnamefont {Ressouche}},\ and\ \bibinfo {author} {\bibfnamefont {B.}~\bibnamefont {F{\aa}k}},\ }\bibfield  {title} {\bibinfo {title} {Kagome ice state in the dipolar spin ice $\ch{Dy2Ti2O7}$},\ }\href@noop {} {\bibfield  {journal} {\bibinfo  {journal} {Physical Review Letters}\ }\textbf {\bibinfo {volume} {97}},\ \bibinfo {pages} {257205} (\bibinfo {year} {2006})}\BibitemShut {NoStop}%
\bibitem [{\citenamefont {Jaubert}\ \emph {et~al.}(2013)\citenamefont {Jaubert}, \citenamefont {Harris}, \citenamefont {Fennell}, \citenamefont {Melko}, \citenamefont {Bramwell},\ and\ \citenamefont {Holdsworth}}]{jaubert2013topological}%
  \BibitemOpen
  \bibfield  {author} {\bibinfo {author} {\bibfnamefont {L.~D.}\ \bibnamefont {Jaubert}}, \bibinfo {author} {\bibfnamefont {M.~J.}\ \bibnamefont {Harris}}, \bibinfo {author} {\bibfnamefont {T.}~\bibnamefont {Fennell}}, \bibinfo {author} {\bibfnamefont {R.~G.}\ \bibnamefont {Melko}}, \bibinfo {author} {\bibfnamefont {S.~T.}\ \bibnamefont {Bramwell}},\ and\ \bibinfo {author} {\bibfnamefont {P.~C.}\ \bibnamefont {Holdsworth}},\ }\bibfield  {title} {\bibinfo {title} {Topological-sector fluctuations and curie-law crossover in spin ice},\ }\href@noop {} {\bibfield  {journal} {\bibinfo  {journal} {Physical Review X}\ }\textbf {\bibinfo {volume} {3}},\ \bibinfo {pages} {011014} (\bibinfo {year} {2013})}\BibitemShut {NoStop}%
\bibitem [{\citenamefont {Chu}\ \emph {et~al.}(2019)\citenamefont {Chu}, \citenamefont {Liu}, \citenamefont {Zhang}, \citenamefont {Noordhoek}, \citenamefont {Riggs}, \citenamefont {Shapiro}, \citenamefont {Serro}, \citenamefont {Yi}, \citenamefont {Mellisa}, \citenamefont {Suresha} \emph {et~al.}}]{chu2019}%
  \BibitemOpen
  \bibfield  {author} {\bibinfo {author} {\bibfnamefont {J.-H.}\ \bibnamefont {Chu}}, \bibinfo {author} {\bibfnamefont {J.}~\bibnamefont {Liu}}, \bibinfo {author} {\bibfnamefont {H.}~\bibnamefont {Zhang}}, \bibinfo {author} {\bibfnamefont {K.}~\bibnamefont {Noordhoek}}, \bibinfo {author} {\bibfnamefont {S.~C.}\ \bibnamefont {Riggs}}, \bibinfo {author} {\bibfnamefont {M.}~\bibnamefont {Shapiro}}, \bibinfo {author} {\bibfnamefont {C.~R.}\ \bibnamefont {Serro}}, \bibinfo {author} {\bibfnamefont {D.}~\bibnamefont {Yi}}, \bibinfo {author} {\bibfnamefont {M.}~\bibnamefont {Mellisa}}, \bibinfo {author} {\bibfnamefont {S.}~\bibnamefont {Suresha}}, \emph {et~al.},\ }\bibfield  {title} {\bibinfo {title} {Possible scale invariant linear magnetoresistance in pyrochlore iridates \ch{Bi2Ir2O7}},\ }\href@noop {} {\bibfield  {journal} {\bibinfo  {journal} {New Journal of Physics}\ }\textbf {\bibinfo {volume} {21}},\ \bibinfo {pages} {113041} (\bibinfo {year} {2019})}\BibitemShut {NoStop}%
\bibitem [{sup()}]{supplementary}%
  \BibitemOpen
  \href@noop {} {\bibinfo {title} {See {S}upplementary {M}aterial at [] for sample synthesis and information; {X}-ray characterization; transmission electron microscopes; resistivity and magnetoresistance measurements; {AC} susceptibility for {YbTO} and {YbTO}-ns; powder {X}-ray diffraction patterns for {YbTO} and {YbTO}-ns; ${T}^{-1}$ dependence of the resistivity for {BIO}/{DTO} at zero field; logarithm of calculated relaxation time as a function of logarithm of temperature; $\mathrm{{B}^{-\frac{1}{2}}}$ dependence of the resistivity of {BIO}/{YTO} at 30 m{K}; reproducibility of the heterostructure {BIO}/{YbTO}; temperature-dependent resistivity from 300 {K} to 30 m{K}.}}\BibitemShut {Stop}%
\bibitem [{\citenamefont {Qi}\ \emph {et~al.}(2012)\citenamefont {Qi}, \citenamefont {Korneta}, \citenamefont {Wan}, \citenamefont {DeLong}, \citenamefont {Schlottmann},\ and\ \citenamefont {Cao}}]{qi2012strong}%
  \BibitemOpen
  \bibfield  {author} {\bibinfo {author} {\bibfnamefont {T.}~\bibnamefont {Qi}}, \bibinfo {author} {\bibfnamefont {O.}~\bibnamefont {Korneta}}, \bibinfo {author} {\bibfnamefont {X.}~\bibnamefont {Wan}}, \bibinfo {author} {\bibfnamefont {L.}~\bibnamefont {DeLong}}, \bibinfo {author} {\bibfnamefont {P.}~\bibnamefont {Schlottmann}},\ and\ \bibinfo {author} {\bibfnamefont {G.}~\bibnamefont {Cao}},\ }\bibfield  {title} {\bibinfo {title} {Strong magnetic instability in correlated metallic \ch{Bi2Ir2O7}},\ }\href@noop {} {\bibfield  {journal} {\bibinfo  {journal} {Journal of Physics: Condensed Matter}\ }\textbf {\bibinfo {volume} {24}},\ \bibinfo {pages} {345601} (\bibinfo {year} {2012})}\BibitemShut {NoStop}%
\bibitem [{\citenamefont {Balberg}(1977)}]{balberg1977critical}%
  \BibitemOpen
  \bibfield  {author} {\bibinfo {author} {\bibfnamefont {I.}~\bibnamefont {Balberg}},\ }\bibfield  {title} {\bibinfo {title} {Critical resistance and magnetoresistance in magnetic materials},\ }\href@noop {} {\bibfield  {journal} {\bibinfo  {journal} {Physica B+ C}\ }\textbf {\bibinfo {volume} {91}},\ \bibinfo {pages} {71} (\bibinfo {year} {1977})}\BibitemShut {NoStop}%
\bibitem [{\citenamefont {Wang}\ \emph {et~al.}(2016)\citenamefont {Wang}, \citenamefont {Barros}, \citenamefont {Chern}, \citenamefont {Maslov},\ and\ \citenamefont {Batista}}]{Batista2016prl}%
  \BibitemOpen
  \bibfield  {author} {\bibinfo {author} {\bibfnamefont {Z.}~\bibnamefont {Wang}}, \bibinfo {author} {\bibfnamefont {K.}~\bibnamefont {Barros}}, \bibinfo {author} {\bibfnamefont {G.-W.}\ \bibnamefont {Chern}}, \bibinfo {author} {\bibfnamefont {D.~L.}\ \bibnamefont {Maslov}},\ and\ \bibinfo {author} {\bibfnamefont {C.~D.}\ \bibnamefont {Batista}},\ }\bibfield  {title} {\bibinfo {title} {Resistivity minimum in highly frustrated itinerant magnets},\ }\href@noop {} {\bibfield  {journal} {\bibinfo  {journal} {Physical Review Letters}\ }\textbf {\bibinfo {volume} {117}},\ \bibinfo {pages} {206601} (\bibinfo {year} {2016})}\BibitemShut {NoStop}%
\bibitem [{\citenamefont {Yan}\ \emph {et~al.}(2017)\citenamefont {Yan}, \citenamefont {Benton}, \citenamefont {Jaubert},\ and\ \citenamefont {Shannon}}]{Yan2017}%
  \BibitemOpen
  \bibfield  {author} {\bibinfo {author} {\bibfnamefont {H.}~\bibnamefont {Yan}}, \bibinfo {author} {\bibfnamefont {O.}~\bibnamefont {Benton}}, \bibinfo {author} {\bibfnamefont {L.}~\bibnamefont {Jaubert}},\ and\ \bibinfo {author} {\bibfnamefont {N.}~\bibnamefont {Shannon}},\ }\bibfield  {title} {\bibinfo {title} {Theory of multiple-phase competition in pyrochlore magnets with anisotropic exchange with application to \ch{Yb2Ti2O7}, \ch{Er2Ti2O7}, and \ch{Er2Sn2O7}},\ }\href@noop {} {\bibfield  {journal} {\bibinfo  {journal} {Physical Review B}\ }\textbf {\bibinfo {volume} {95}},\ \bibinfo {pages} {094422} (\bibinfo {year} {2017})}\BibitemShut {NoStop}%
\bibitem [{\citenamefont {Fuchs}\ \emph {et~al.}(2005)\citenamefont {Fuchs}, \citenamefont {Gangardt}, \citenamefont {Keilmann},\ and\ \citenamefont {Shlyapnikov}}]{fuchs2005spin}%
  \BibitemOpen
  \bibfield  {author} {\bibinfo {author} {\bibfnamefont {J.}~\bibnamefont {Fuchs}}, \bibinfo {author} {\bibfnamefont {D.}~\bibnamefont {Gangardt}}, \bibinfo {author} {\bibfnamefont {T.}~\bibnamefont {Keilmann}},\ and\ \bibinfo {author} {\bibfnamefont {G.}~\bibnamefont {Shlyapnikov}},\ }\bibfield  {title} {\bibinfo {title} {Spin waves in a one-dimensional spinor bose gas},\ }\href@noop {} {\bibfield  {journal} {\bibinfo  {journal} {Physical Review Letters}\ }\textbf {\bibinfo {volume} {95}},\ \bibinfo {pages} {150402} (\bibinfo {year} {2005})}\BibitemShut {NoStop}%
\bibitem [{\citenamefont {Fukazawa}\ \emph {et~al.}(2002)\citenamefont {Fukazawa}, \citenamefont {Melko}, \citenamefont {Higashinaka}, \citenamefont {Maeno},\ and\ \citenamefont {Gingras}}]{fukazawa2002magnetic}%
  \BibitemOpen
  \bibfield  {author} {\bibinfo {author} {\bibfnamefont {H.}~\bibnamefont {Fukazawa}}, \bibinfo {author} {\bibfnamefont {R.}~\bibnamefont {Melko}}, \bibinfo {author} {\bibfnamefont {R.}~\bibnamefont {Higashinaka}}, \bibinfo {author} {\bibfnamefont {Y.}~\bibnamefont {Maeno}},\ and\ \bibinfo {author} {\bibfnamefont {M.}~\bibnamefont {Gingras}},\ }\bibfield  {title} {\bibinfo {title} {Magnetic anisotropy of the spin-ice compound \ch{Dy2Ti2O7}},\ }\href@noop {} {\bibfield  {journal} {\bibinfo  {journal} {Physical Review B}\ }\textbf {\bibinfo {volume} {65}},\ \bibinfo {pages} {054410} (\bibinfo {year} {2002})}\BibitemShut {NoStop}%
\bibitem [{\citenamefont {Zhang}\ \emph {et~al.}(2019)\citenamefont {Zhang}, \citenamefont {Changlani}, \citenamefont {Plumb}, \citenamefont {Tchernyshyov},\ and\ \citenamefont {Moessner}}]{zhang2019}%
  \BibitemOpen
  \bibfield  {author} {\bibinfo {author} {\bibfnamefont {S.}~\bibnamefont {Zhang}}, \bibinfo {author} {\bibfnamefont {H.~J.}\ \bibnamefont {Changlani}}, \bibinfo {author} {\bibfnamefont {K.~W.}\ \bibnamefont {Plumb}}, \bibinfo {author} {\bibfnamefont {O.}~\bibnamefont {Tchernyshyov}},\ and\ \bibinfo {author} {\bibfnamefont {R.}~\bibnamefont {Moessner}},\ }\bibfield  {title} {\bibinfo {title} {Dynamical structure factor of the three-dimensional quantum spin liquid candidate \ch{NaCaNi2F7}},\ }\href@noop {} {\bibfield  {journal} {\bibinfo  {journal} {Physical Review Letters}\ }\textbf {\bibinfo {volume} {122}},\ \bibinfo {pages} {167203} (\bibinfo {year} {2019})}\BibitemShut {NoStop}%
\bibitem [{\citenamefont {Laurell}\ \emph {et~al.}(2021)\citenamefont {Laurell}, \citenamefont {Scheie}, \citenamefont {Mukherjee}, \citenamefont {Koza}, \citenamefont {Enderle}, \citenamefont {Tylczynski}, \citenamefont {Okamoto}, \citenamefont {Coldea}, \citenamefont {Tennant},\ and\ \citenamefont {Alvarez}}]{Laurell2021}%
  \BibitemOpen
  \bibfield  {author} {\bibinfo {author} {\bibfnamefont {P.}~\bibnamefont {Laurell}}, \bibinfo {author} {\bibfnamefont {A.}~\bibnamefont {Scheie}}, \bibinfo {author} {\bibfnamefont {C.~J.}\ \bibnamefont {Mukherjee}}, \bibinfo {author} {\bibfnamefont {M.~M.}\ \bibnamefont {Koza}}, \bibinfo {author} {\bibfnamefont {M.}~\bibnamefont {Enderle}}, \bibinfo {author} {\bibfnamefont {Z.}~\bibnamefont {Tylczynski}}, \bibinfo {author} {\bibfnamefont {S.}~\bibnamefont {Okamoto}}, \bibinfo {author} {\bibfnamefont {R.}~\bibnamefont {Coldea}}, \bibinfo {author} {\bibfnamefont {D.~A.}\ \bibnamefont {Tennant}},\ and\ \bibinfo {author} {\bibfnamefont {G.}~\bibnamefont {Alvarez}},\ }\bibfield  {title} {\bibinfo {title} {Quantifying and controlling entanglement in the quantum magnet \ch{Cs2CoCl4}},\ }\href@noop {} {\bibfield  {journal} {\bibinfo  {journal} {Physical Review Letters}\ }\textbf {\bibinfo {volume} {127}},\ \bibinfo {pages} {037201} (\bibinfo {year} {2021})}\BibitemShut {NoStop}%
\bibitem [{\citenamefont {Hirschberger}\ \emph {et~al.}(2019)\citenamefont {Hirschberger}, \citenamefont {Czajka}, \citenamefont {Koohpayeh}, \citenamefont {Wang},\ and\ \citenamefont {Ong}}]{hirschberger2019enhanced}%
  \BibitemOpen
  \bibfield  {author} {\bibinfo {author} {\bibfnamefont {M.}~\bibnamefont {Hirschberger}}, \bibinfo {author} {\bibfnamefont {P.}~\bibnamefont {Czajka}}, \bibinfo {author} {\bibfnamefont {S.}~\bibnamefont {Koohpayeh}}, \bibinfo {author} {\bibfnamefont {W.}~\bibnamefont {Wang}},\ and\ \bibinfo {author} {\bibfnamefont {N.~P.}\ \bibnamefont {Ong}},\ }\bibfield  {title} {\bibinfo {title} {Enhanced thermal hall conductivity below 1 kelvin in the pyrochlore magnet \ch{Yb2Ti2O7}},\ }\href@noop {} {\bibfield  {journal} {\bibinfo  {journal} {arXiv preprint arXiv:1903.00595}\ } (\bibinfo {year} {2019})}\BibitemShut {NoStop}%
\bibitem [{\citenamefont {Chatterjee}\ \emph {et~al.}(2019)\citenamefont {Chatterjee}, \citenamefont {Rodriguez-Nieva},\ and\ \citenamefont {Demler}}]{chatterjee2019}%
  \BibitemOpen
  \bibfield  {author} {\bibinfo {author} {\bibfnamefont {S.}~\bibnamefont {Chatterjee}}, \bibinfo {author} {\bibfnamefont {J.~F.}\ \bibnamefont {Rodriguez-Nieva}},\ and\ \bibinfo {author} {\bibfnamefont {E.}~\bibnamefont {Demler}},\ }\bibfield  {title} {\bibinfo {title} {Diagnosing phases of magnetic insulators via noise magnetometry with spin qubits},\ }\href@noop {} {\bibfield  {journal} {\bibinfo  {journal} {Physical Review B}\ }\textbf {\bibinfo {volume} {99}},\ \bibinfo {pages} {104425} (\bibinfo {year} {2019})}\BibitemShut {NoStop}%
\bibitem [{\citenamefont {Li}\ \emph {et~al.}(2013)\citenamefont {Li}, \citenamefont {Xu}, \citenamefont {Fan}, \citenamefont {Zhang}, \citenamefont {Lv}, \citenamefont {Ni}, \citenamefont {Zhao},\ and\ \citenamefont {Sun}}]{li2013single}%
  \BibitemOpen
  \bibfield  {author} {\bibinfo {author} {\bibfnamefont {Q.}~\bibnamefont {Li}}, \bibinfo {author} {\bibfnamefont {L.}~\bibnamefont {Xu}}, \bibinfo {author} {\bibfnamefont {C.}~\bibnamefont {Fan}}, \bibinfo {author} {\bibfnamefont {F.}~\bibnamefont {Zhang}}, \bibinfo {author} {\bibfnamefont {Y.}~\bibnamefont {Lv}}, \bibinfo {author} {\bibfnamefont {B.}~\bibnamefont {Ni}}, \bibinfo {author} {\bibfnamefont {Z.}~\bibnamefont {Zhao}},\ and\ \bibinfo {author} {\bibfnamefont {X.}~\bibnamefont {Sun}},\ }\bibfield  {title} {\bibinfo {title} {Single crystal growth of the pyrochlores \ch{R2Ti2O7} (\ch{R}= rare earth) by the optical floating-zone method},\ }\href@noop {} {\bibfield  {journal} {\bibinfo  {journal} {Journal of crystal growth}\ }\textbf {\bibinfo {volume} {377}},\ \bibinfo {pages} {96} (\bibinfo {year} {2013})}\BibitemShut {NoStop}%
\bibitem [{\citenamefont {Yang}\ \emph {et~al.}(2017)\citenamefont {Yang}, \citenamefont {Xie}, \citenamefont {Zhu}, \citenamefont {Park}, \citenamefont {Chen}, \citenamefont {Losovyj}, \citenamefont {Li}, \citenamefont {Liu}, \citenamefont {Starr}, \citenamefont {Acosta} \emph {et~al.}}]{yang2017epitaxial}%
  \BibitemOpen
  \bibfield  {author} {\bibinfo {author} {\bibfnamefont {W.}~\bibnamefont {Yang}}, \bibinfo {author} {\bibfnamefont {Y.}~\bibnamefont {Xie}}, \bibinfo {author} {\bibfnamefont {W.}~\bibnamefont {Zhu}}, \bibinfo {author} {\bibfnamefont {K.}~\bibnamefont {Park}}, \bibinfo {author} {\bibfnamefont {A.}~\bibnamefont {Chen}}, \bibinfo {author} {\bibfnamefont {Y.}~\bibnamefont {Losovyj}}, \bibinfo {author} {\bibfnamefont {Z.}~\bibnamefont {Li}}, \bibinfo {author} {\bibfnamefont {H.}~\bibnamefont {Liu}}, \bibinfo {author} {\bibfnamefont {M.}~\bibnamefont {Starr}}, \bibinfo {author} {\bibfnamefont {J.~A.}\ \bibnamefont {Acosta}}, \emph {et~al.},\ }\bibfield  {title} {\bibinfo {title} {Epitaxial thin films of pyrochlore iridate $\ch{Bi}_{2+x}\ch{Ir}_{2-y}\ch{O}_{7-\delta}$: Structure, defects and transport properties},\ }\href@noop {} {\bibfield  {journal} {\bibinfo  {journal} {Scientific Reports}\ }\textbf {\bibinfo {volume} {7}},\ \bibinfo {pages} {7740} (\bibinfo {year} {2017})}\BibitemShut {NoStop}%
\end{thebibliography}%
	
\end{document}